\documentclass[referee]{aa}
\usepackage{graphicx}
\usepackage{txfonts}

\usepackage{subfig}
\usepackage{float}
\usepackage{comment}
\usepackage{caption}
\usepackage{bm}
\usepackage{amsfonts}
\usepackage{amssymb}
\usepackage{array}
\usepackage{natbib}

\usepackage{url}

\usepackage{color}
\usepackage{multirow}

\usepackage[ruled,vlined]{algorithm2e}

\begin{document}

\title{Visibility Interpolation in Solar Hard X-ray Imaging: Application to {\em{RHESSI}} and {\em{STIX}}}

\author{Emma Perracchione\inst{1} \and Paolo Massa\inst{1}  \and Anna Maria Massone\inst{1,2} \and Michele Piana\inst{1,2}}

\institute{Dipartimento di Matematica,  Universit\`{a} degli Studi di Genova, Via   Dodecaneso 35, 16146 Genova, Italy\\
\and
CNR-SPIN, Via Dodecaneso 33, 16146 Genova, Italy\\
\email{perracchione@dima.unige.it}, \email{massa.p@dima.unige.it}, \email{massone@dima.unige.it},
\email{piana@dima.unige.it}}


\abstract
{}
{Space telescopes for solar hard X-ray imaging provide observations made of sampled Fourier components of the incoming photon flux. The aim of this study is to design an image reconstruction method relying on enhanced visibility interpolation in the Fourier domain.}
{The interpolation-based method is applied on synthetic visibilities generated by means of the simulation software implemented within the framework of the \emph{Spectrometer/Telescope for Imaging X-rays (STIX)} mission on board {\em{Solar Orbiter}}. An application to experimental visibilities observed by the {\emph{Reuven Ramaty High Energy Solar Spectroscopic Imager (RHESSI)}} is also considered. In order to interpolate these visibility data we have utilized an approach based on Variably Scaled Kernels (VSKs), which are able to realize feature augmentation by exploiting prior information on the flaring source and which are used here, for the first time, for image reconstruction purposes.}
{When compared to an interpolation-based reconstruction algorithm previously introduced for {\emph{RHESSI}}, VSKs offer significantly better performances, particularly in the case of {\emph{STIX}} imaging, which is characterized by a notably sparse sampling of the Fourier domain. In the case of {\emph{RHESSI}} data, this novel approach is particularly reliable when either the flaring sources are characterized by narrow, ribbon-like shapes or high-resolution detectors are utilized for observations.}
{The use of VSKs for interpolating hard X-ray visibilities allows a notable image reconstruction accuracy when the information on the flaring source is encoded by a small set of scattered Fourier data and when the visibility surface is affected by significant oscillations in the frequency domain.}


\titlerunning{Visibility interpolation}
\authorrunning{Perracchione et al}

\maketitle



\date{\today}


\section{Introduction}

\begin{figure}
\centering
    \includegraphics[scale=0.2]{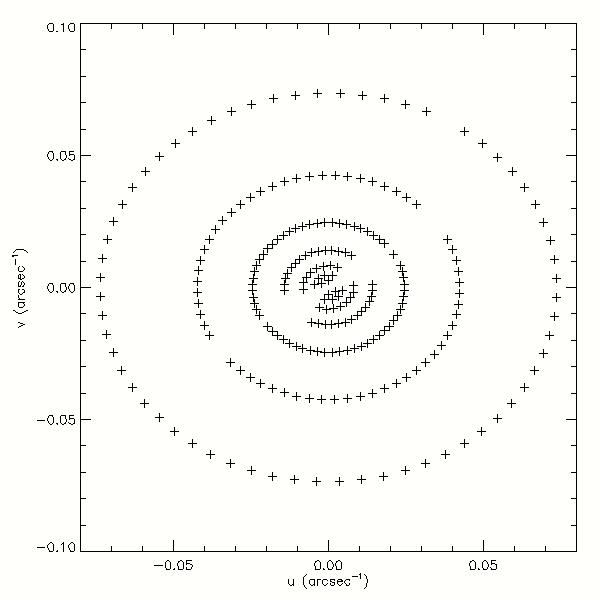} \hskip 1.2cm
    \includegraphics[scale=0.2]{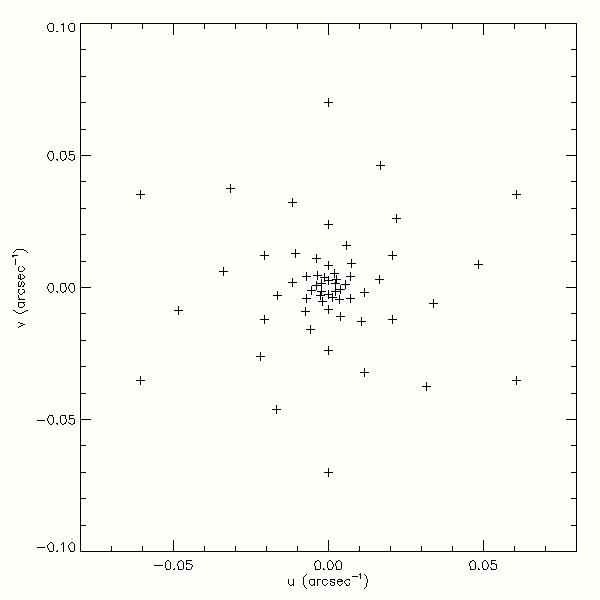}
    \caption{The sampling of the $(u,v)$ plane provided by {\em{RHESSI}} (left panel) and {\em{STIX}} (right panel).}
    \label{figure:fig-1}
\end{figure}

The use of Fourier methods in astronomical imaging is mainly related to radio interferometry \citep{richard2017interferometry}. However, in the last three decades, this approach has been utilized also in the case of solar hard X-ray telescopes that have been conceived in order to provide spatial Fourier components of the photon flux emitted via either bremsstrahlung or thermal processes during solar flares \citep{enlighten1658,krucker2020spectrometer}. These Fourier components, named {\em{visibilities}}, are sampled by the hard X-ray instrument in the two dimensional Fourier space, named $(u,{\mbox{v}})$-plane, in a sparse way, according to a geometry depending on the instrument design. By instance, the {\em{Reuven Ramaty High Energy Spectroscopic Imager (RHESSI)}} relies on the use of a set of nine rotating modulation collimators (RMCs) whose Full Width at Half Maximum (FWHM) is logarithmically spaced between $2.3$ and $183$ arcsec \citep{2002SoPh}. Each RMC measures visibilities on a circle of points in the $(u,v)$-space with a spatial frequency that corresponds to its angular resolution and a position angle that varies according to the spacecraft rotation (see Figure \ref{figure:fig-1}, left panel). On the other hand, the {\em{Spectrometer/Telescope for Imaging X-rays (STIX)}} on-board {\em{Solar Orbiter}} is based on the Moir\'e pattern technology \citep{STIX1,2019A&A...624A.130M} and its $30$ collimators sample the $(u,{\mbox{v}})$-plane over a set of six spirals for a FWHM resolution coarser than $7$ arcsec (see Figure \ref{figure:fig-1}, right panel). 

Image reconstruction methods in solar hard X-ray astronomy rely on procedures that allow some sort of interpolation/extrapolation in the $(u,{\mbox{v}})$-space in order to recover information in between the sampled frequencies, for reducing the imaging artifacts and, outside the sampling domain, for obtaining super-resolution effects. Most methods accomplish these objectives by imposing constraints in the image domain, either by optimizing parameters associated to predefined image shapes via comparison with observations \citep{Aschwanden,sciacchitano2018identification}, or by minimizing regularization functionals that combine a fitting term with a stability term \citep{felix2017compressed,duval2018solar,massa2020mem_ge}.

However, the most straightforward approach to interpolation/extrapolation in visibility-based imaging is probably the one implemented in the uv$\_$smooth method \citep{009HAMassoneRDXI}, which is inspired by standard gridding approaches utilized in radio-astronomy. In particular, uv$\_$smooth starts from the observation that the coverage of the $(u,{\mbox{v}})$-plane offered by hard X-ray instruments is much sparser than that typical of radio astronomy and therefore utilizes spline interpolation at spatial frequencies smaller than the largest sampled frequencies and soft-thresholding on the image to reduce the ringing effects due to a naive and unconstrained Fourier transform inversion procedure \citep{daubechies2004iterative,Massone1}. This approach can exploit Fast Fourier Transform (FFT) in the inversion process and is characterized by a satisfactory reliability when reconstructing extended sources \citep{guo2013specific,guo2012determination,guo2012properties,caspi2015hard}; however, several applications \citep{dennis2019remarkably,bonettini2014accelerated} showed that uv$\_$smooth does not work properly when it is applied to visibility sets characterized by significant oscillations in the $(u,{\mbox{v}})$-plane. This misbehavior is essentially due to the fact that the interpolation algorithm utilized in uv$\_$smooth is not optimal and often misses the oscillating frequency information related to very narrow or well-separated sources (or, in the case of {\em{RHESSI}}, associated to the use of detectors with fine grids in the observation process).

The present paper proposes an enhanced release of uv$\_$smooth, based on the use of an advanced approach to interpolation in the frequency domain. Specifically, this approach relies on the use of Variably Scaled Kernels (VSKs), which are able to include {\em{a priori}} information in the interpolation process \citep{Bozzini1,vskmpi}. This additional knowledge is implicitly put into the kernel via a {\em{scaling function}} that determines the accuracy of the approximation process and that is linked to a first coarse reconstruction of the sought image. As far as the practical implementation of the VSK setting is concerned, in this study we considered the Mat\'ern $C^0$ kernel, which takes advantage of a low regularity degree and of a better numerical stability \citep{Matern}.

The plan of the paper is as follows. Section 2 illustrates the interpolation process based on VSKs. Section 3 describes the overall image reconstruction approach relying on the use of interpolation in the $(u,{\mbox{v}})$-plane and of the soft-thresholding technique applied for image reconstruction. Section 4 contains some validation tests performed against both synthetic {\em{STIX}} visibilities and experimental {\em{RHESSI}} observations. Our conclusions are offered in Section 5.

\section{Interpolation in the Fourier domain}
Visibility-based hard X-ray telescopes provide experimental measurements of the Fourier transform of the incoming photon flux at specific points of the spatial frequency plane. We denote with ${\bf{f}}$ the vector whose components are the discretized values of the incoming flux, with ${\bf{F}}$ the discretized Fourier transform, with $ \{ {\bf u}_i=(u_i,v_i) \}_{i=1}^{n}$ the set of sampled points in the $(u,{\mbox{v}})$-plane, with ${\bf{V}}$ the vector whose $n$ components are the observed visibilities and with $\chi$ the binary mask returning $1$ at frequencies $\{{\bf{u}}_i\}_{i=1}^{n}$ and zero elsewhere. Then, the image formation model in this framework can be approximated by
\begin{equation}\label{b1}
{\bf{V}} = \chi \cdot {\bf{F}} {\bf{f}}~,
\end{equation}
where the symbol $\cdot$ denotes the entry-wise product. The uv$\_$smooth code incorporated in the SSW tree and validated in the case of {\em{RHESSI}} visibilities addresses equation (\ref{b1}) by means of an interpolation/extrapolation procedure in which the interpolation step is carried out via an algorithm based on spline functions and the extrapolation step is realized by means of a soft-thresholding scheme \citep{daubechies2004iterative,Massone1}. In the present paper we want to generalize the interpolation step of uv$\_$smooth by means of a more sophisticated numerical technique, in order to improve uv$\_$smooth performances, particularly in the case when visibility oscillations are significant. 

In general, any interpolation approach seeks for a function, namely $P$, that matches the given measurements at their corresponding locations. Thus an interpolant of the visibilities is constructed in such a way that 
\begin{equation}\label{b1-1}
P(\boldsymbol{u}_i)=\boldsymbol{V}_i, \quad i=1,\ldots,n.    
\end{equation}
Typically, any interpolating function is of the form
\begin{equation}\label{b2}
P({\bf{u}}) = \sum_{k=1}^{n} a_k b_k({\bf{u}})~,
\end{equation}
where $\{b_1({\bf{u}}),\ldots,b_n({\bf{u}})\}$ is a set of appropriate basis functions and ${\bf{u}}$ is a vector in the interpolation domain. A possible choice for these basis functions is represented by the so-called Radial Basis Functions (RBFs), see e.g \citep{Fasshauer}, which have the property that
\begin{equation}\label{b3}
b_k({\bf{u}}) = \phi(\|{\bf{u}}-{\bf{u}}_k\|),~~~~k=1,\ldots,n~,
\end{equation}
where $\phi$ is a specific RBF. 
In order to incorporate possible prior information in the interpolation process, the Variably Scaled Kernels (VSKs) represent a specific implementation of RBFs in which
\begin{equation}\label{b4}
b_k({\bf{u}}) = \phi(\|({\bf{u}},\psi({\bf{u}}))- ({\bf{u}}_k,\psi({\bf{u}}_k))\|), 
~~~k=1,\ldots,n~,
\end{equation}
and where $\psi$ is the so-called scaling function encoding such prior information on the emitting source ${\bf{f}}$. Therefore, once the functions $\phi$ and $\psi$ are chosen, by imposing the interpolation conditions (\ref{b1-1}) the interpolation problem is reduced to the solution of the linear system
\begin{equation}\label{b5}
K{\bf{a}} = {\bf{V}}~,
\end{equation}
where ${\bf{a}}=(a_1,\ldots,a_n)^T$, ${\bf{V}}=({\bf{v}}_1,\ldots,{\bf{v}}_n)^T$ and $K_{ij}=\phi(\|({\bf{u}}_i,\psi({\bf{u}}_i))- ({\bf{u}}_j,\psi({\bf{u}}_j)\|)$, $i,j=1,\ldots,n$. Once system (\ref{b5}) is solved, the computed vector ${\bf{a}}$ is used to evaluate the interpolating function $P({\bf{u}})$ on the $N$ points $\{\bar{{\bf{u}}}_1,\ldots,\bar{{\bf{u}}}_N\}$ of a regular mesh of the $(u,{\mbox{v}})$-plane, with $N >> n$. This provides the visibility surface ${\bf{\overline{V}}}$ such that
\begin{equation}\label{bb5}
{\bf{\overline{V}}}_k =  P(\bar{{\bf u}}_k) = \sum_{i=1}^n a_i  \phi(\|({\bf{{\overline{u}}}}_k,\psi({\bf{{\overline{u}}}}_k))- ({\bf{u}}_i,\psi({\bf{u}}_i))\|),
~~~k=1,\ldots,N~.
\end{equation}
Equation (\ref{bb5}) implies that, after interpolation, the reconstruction problem for visibility-based interpolation has become
\begin{equation}\label{bbb5}
{\bf{\overline{V}}} = {\bf{\overline{F}}} {\bf{\overline{f}}}~,
\end{equation}
where ${\bf{\overline{F}}}$ is the $N \times N$ discretized Fourier transform and ${\bf{\overline{f}}}$ is the $N \times 1$ vector to reconstruct.


Two comments are probably relevant in conclusion of this subsection. First, from a technical viewpoint, the choice of $\phi$ and $\psi$ should guarantee numerical stability of system (\ref{b5}). Moreover, at a more general level, VSK approaches map the original measured data into a higher dimension space and therefore can be considered as a feature augmentation strategy. It follows that the definition of the scaling function plays a crucial role for the final outcome of this approach and the idea is to select it so that it mimics the samples as shown in \citep{vskmpi,vskjump,romani}.

\section{Image reconstruction}

The implementation of an image reconstruction process relying on the interpolation procedure described in the previous section needs the definition of a pipeline made of the following steps:
\begin{enumerate}
    \item Construction of the matrix $K$. This step needs the choice of the function $\phi$ generating the RBFs and of the scaling function $\psi$, which implies to account for some prior information on the source image. As far as $\phi$ is concerned, we have chosen the Gaussian-like Mat\'ern function
\begin{equation}\label{b3-1}
\phi(\|{\bf{u}}-{\bf{u}}_k\|) = {\rm e}^{-\|{\bf{u}}-{\bf{u}}_k\|}~.
\end{equation}
As for $\psi$, in this study we have implemented two possible choices, based on coarse estimates of the X-ray source to reconstruct:
    \begin{itemize}
        \item We have applied the inverse Discrete Fourier Transform to the visibility set and used the Fourier projection of the corresponding back-projected map as the scaling function.
        \item We have applied CLEAN to the visibility set and used the Fourier projection of the map of the CLEAN components as the scaling function.
        \end{itemize}
    \item Solution of equation (\ref{b5}). This is a square and rather well-conditioned linear system and therefore standard numerics for computing $K^{-1}$ works properly in the case of input data characterized by large signal-to-noise ratios. When the data statistics is low the system is solved by means of the equally standard Tikhonov method \citep{2003A&A...405..325M}.
    \item Reconstruction of the image ${\bf{f}}$. To this aim we have implemented a soft-thresholding approach based on the projected Landweber iterative scheme \citep{1996JOSAA..13.1516P,piana1997projected}
    \begin{equation}\label{c1}
    {\bf{{\overline{f}}}}^{(k+1)} = {\cal{P}}_+[{\bf{\overline{f}}}^{(k)} + {\bf{\overline{F}}}^T({\bf{\overline{V}}} -
    {\bf{\overline{F}}}{\bf{\overline{f}}}^{(k)})]~,
    \end{equation}
    where ${\cal{P}}_+$ pixel-wise imposes a positivity constraint.
    In the present implementation we have assumed the initialization ${\bf{f}}=0$ and a stopping rule that relies on a check on the $\chi^2$ values \citep{Massone1}.
  \end{enumerate}

The main advantages of this scheme are essentially two. First, the positivity constraint induces super-resolution effects, since it allows extrapolating the frequency information outside the support of the interpolated visibility surface \citep{1996JOSAA..13.1516P}. Second, the implementation of the iterative scheme is made computationally effective by the use of an FFT routine performing the required forward and backward Fourier transformation. We also point out that a well-established weakness of CLEAN is the fact that the determination of the reconstructed CLEAN map from the map of the CLEAN components is typically realized by means of convolution with an idealized point spread function whose FWHM is chosen by means of totally heuristic considerations. Choosing $\psi$ as the map of the CLEAN components is a way to exploit it in a completely objective way, within the framework of an automatic image reconstruction method.

\section{Applications to the reconstruction of flaring sources}
In this section we discuss the effectiveness of this enhanced release of uv$\_$smooth for visibility-based image reconstruction by considering tests on both synthetic simulations obtained by means of the {\em{STIX}} simulation software and experimental {\em{RHESSI}} observations.

\subsection{STIX simulated visibilities}

\begin{figure}[h!] 
    \centering
  \includegraphics[scale=0.11]{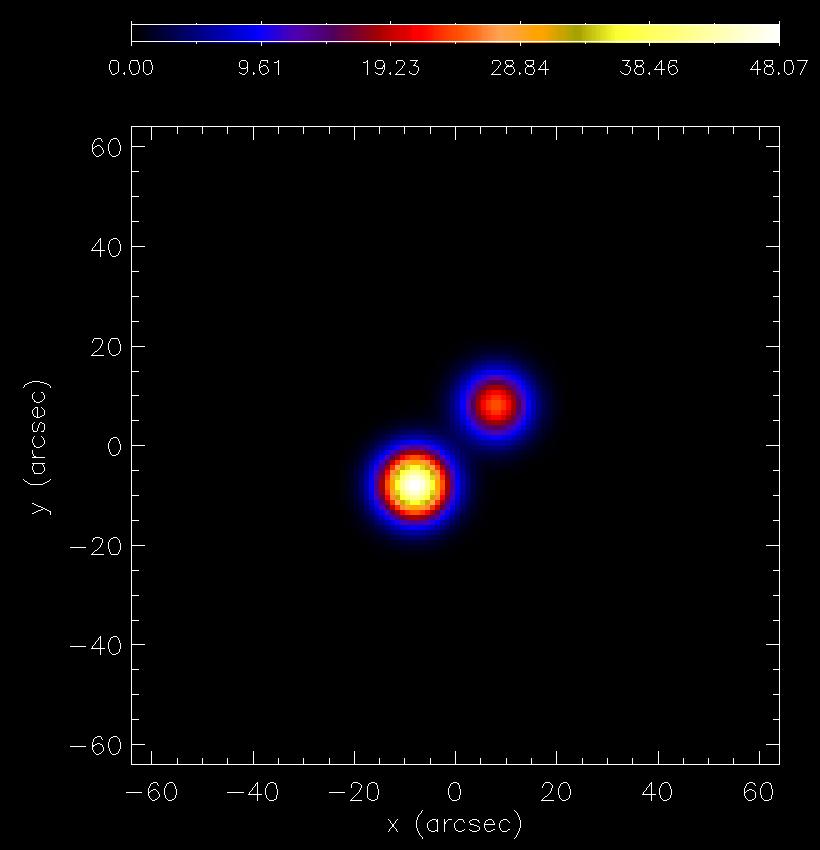}
  \includegraphics[scale=0.11]{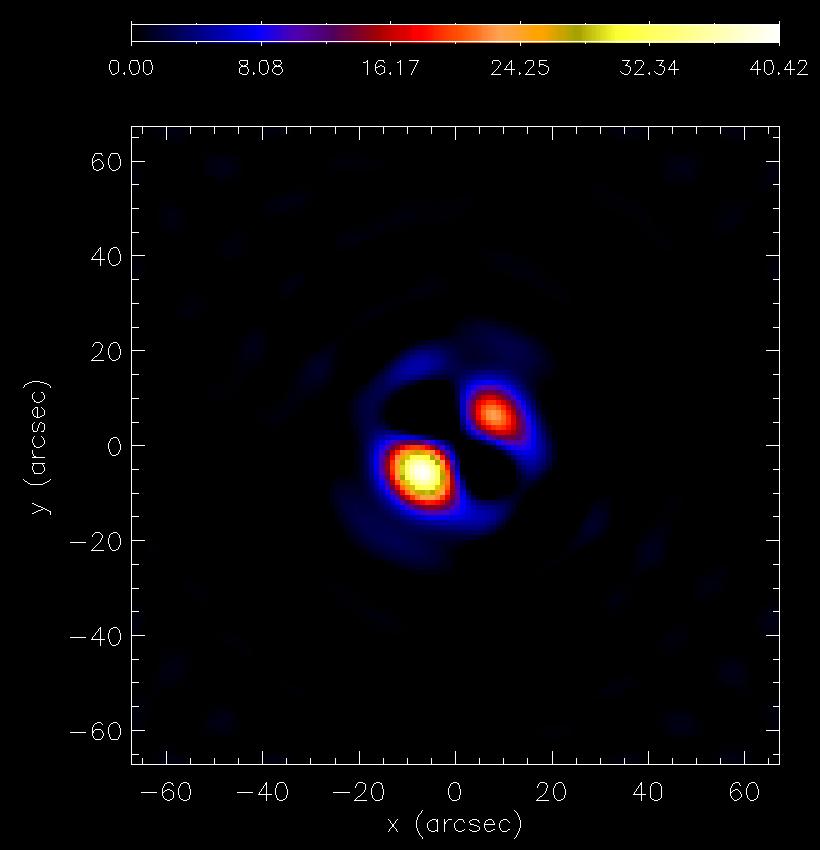}
  \includegraphics[scale=0.11]{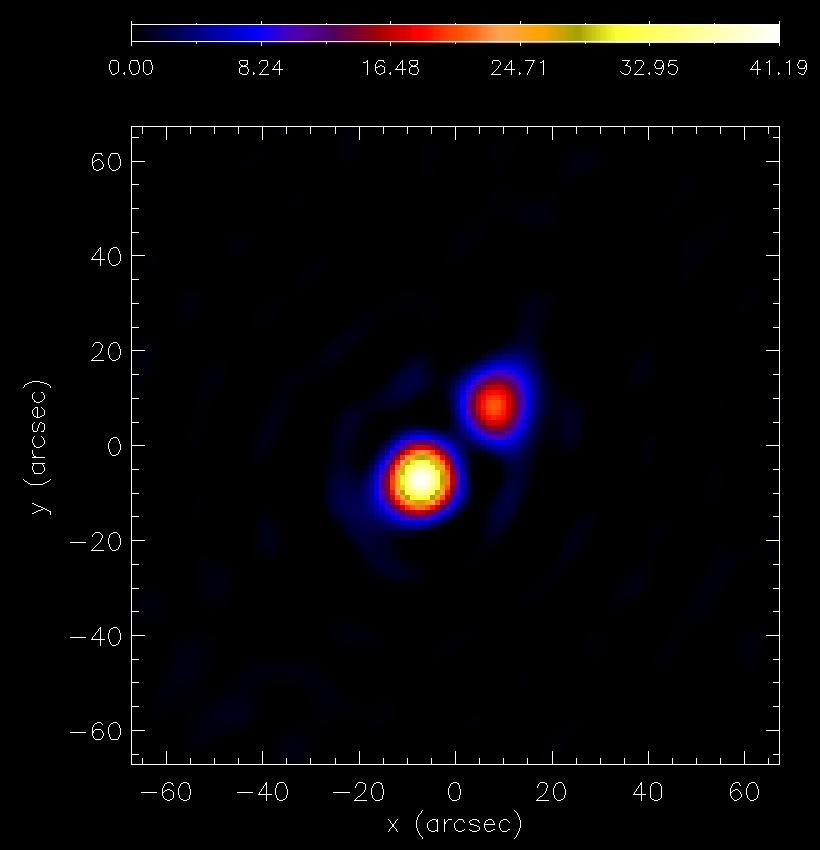}
  \includegraphics[scale=0.11]{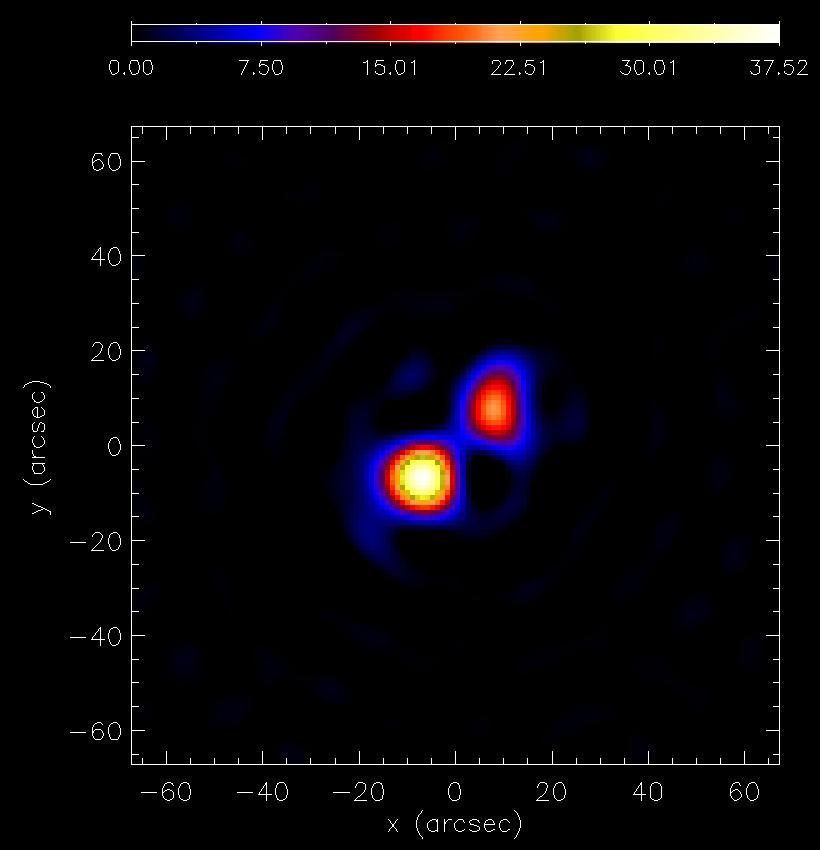}\vskip 0.1cm
   \includegraphics[scale=0.11]{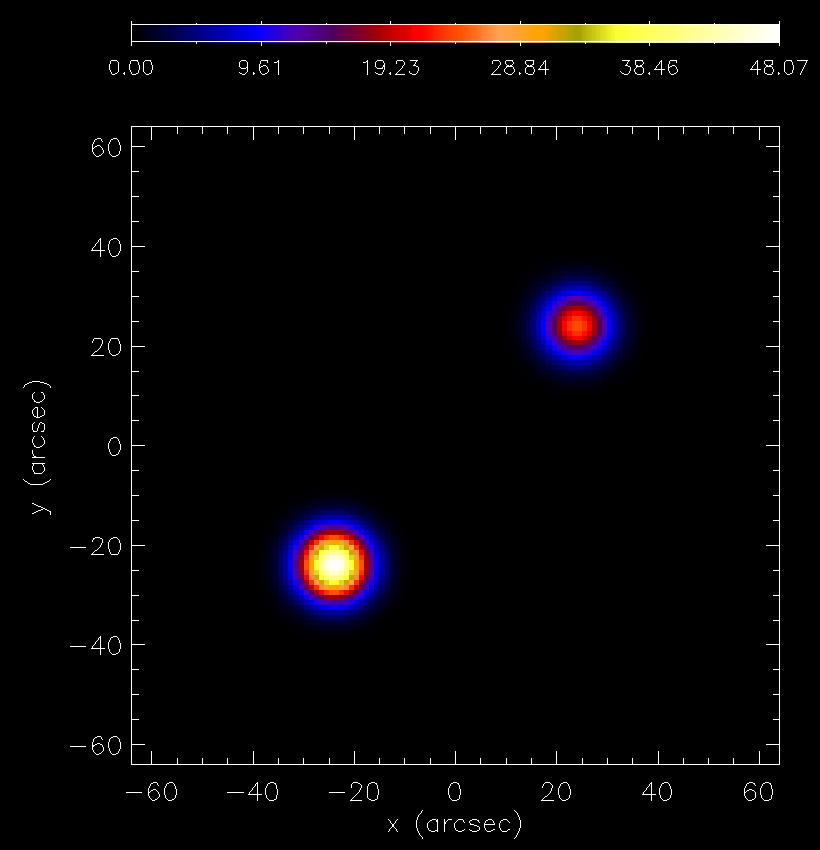}
  \includegraphics[scale=0.11]{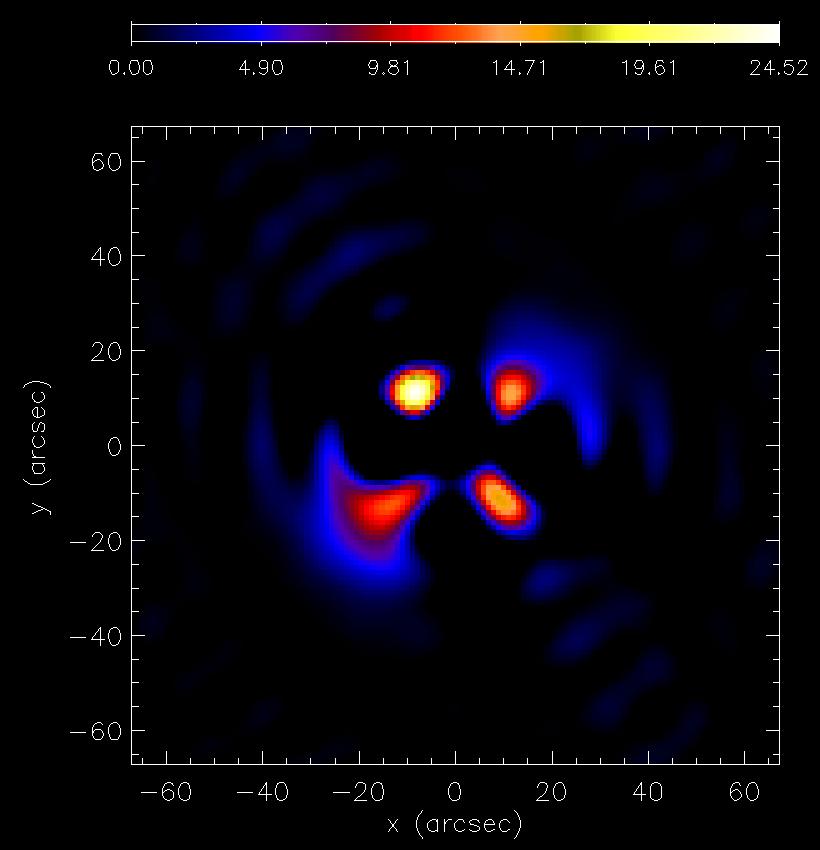}
  \includegraphics[scale=0.11]{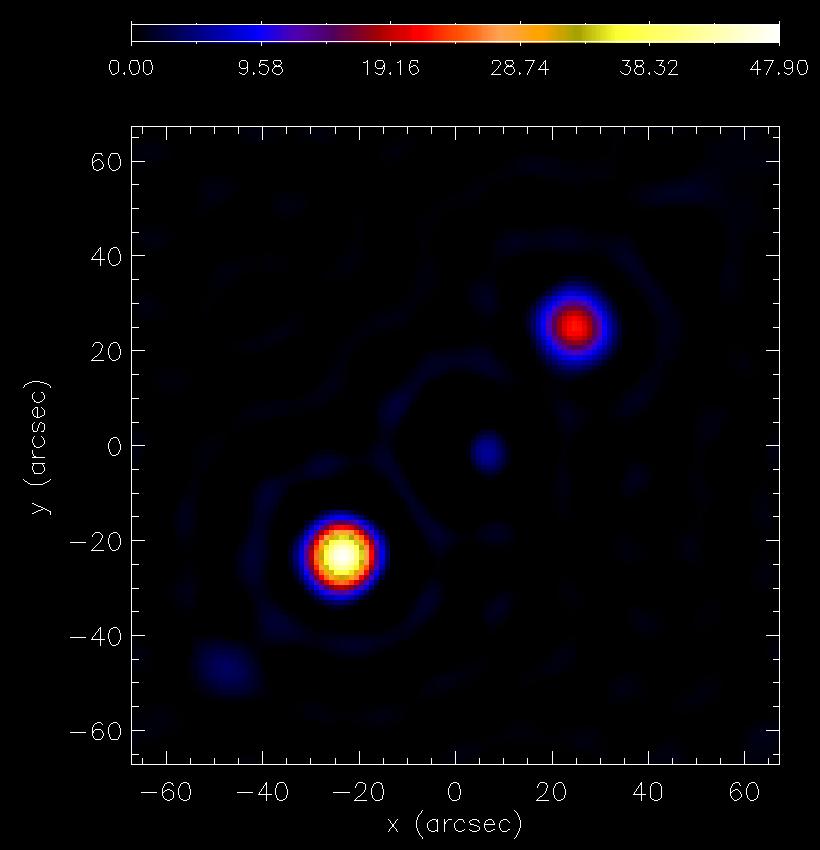}
  \includegraphics[scale=0.11]{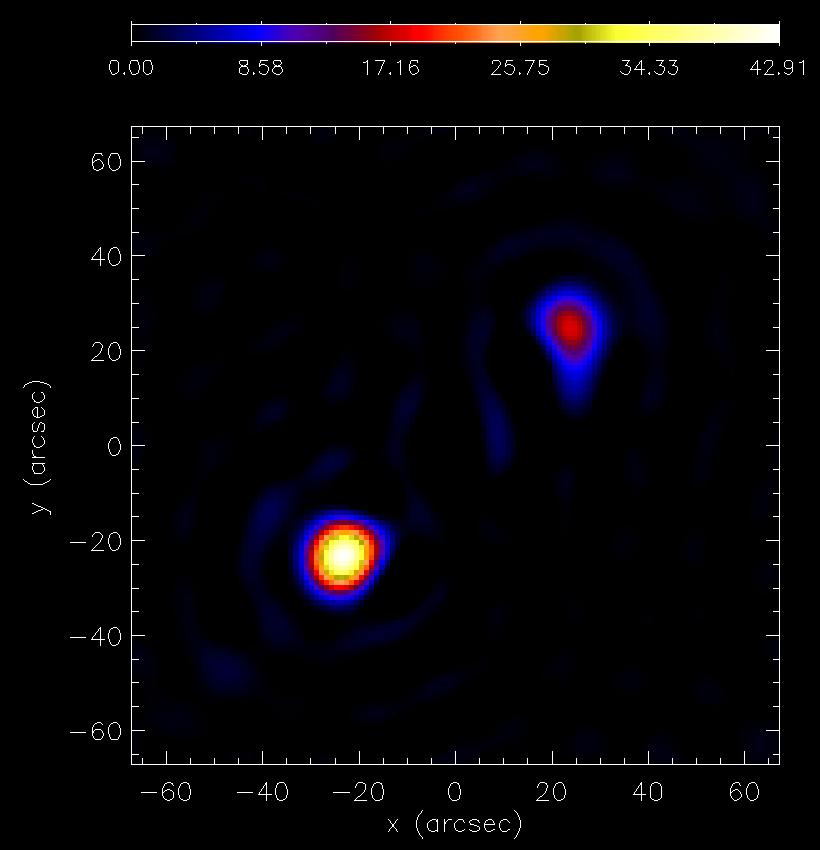}  \vskip 0.1cm   
   \includegraphics[scale=0.11]{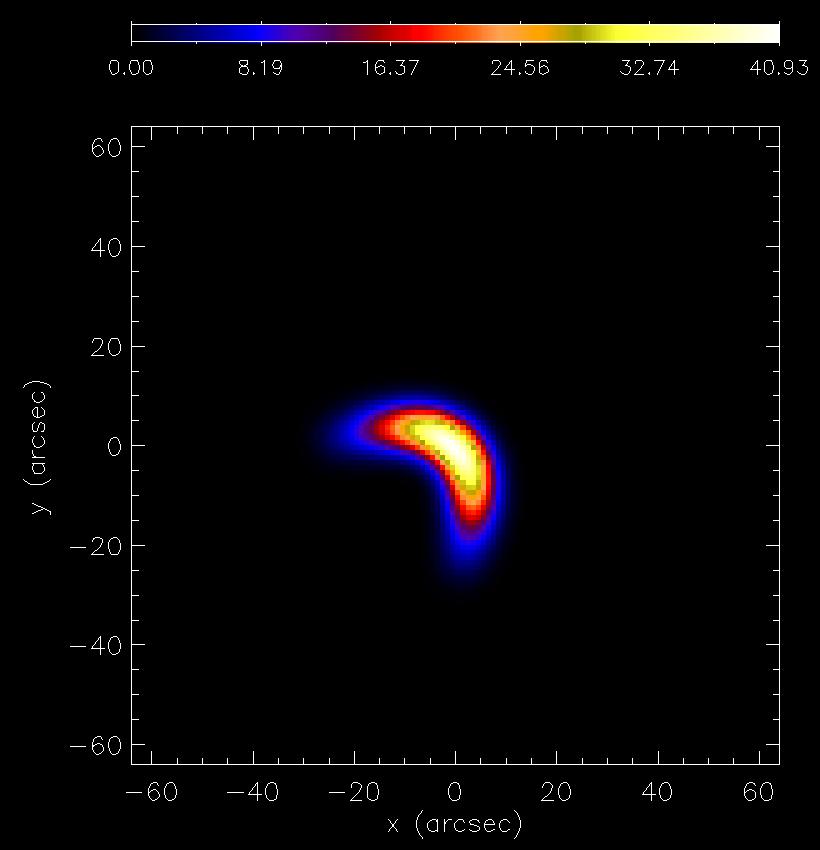}
  \includegraphics[scale=0.11]{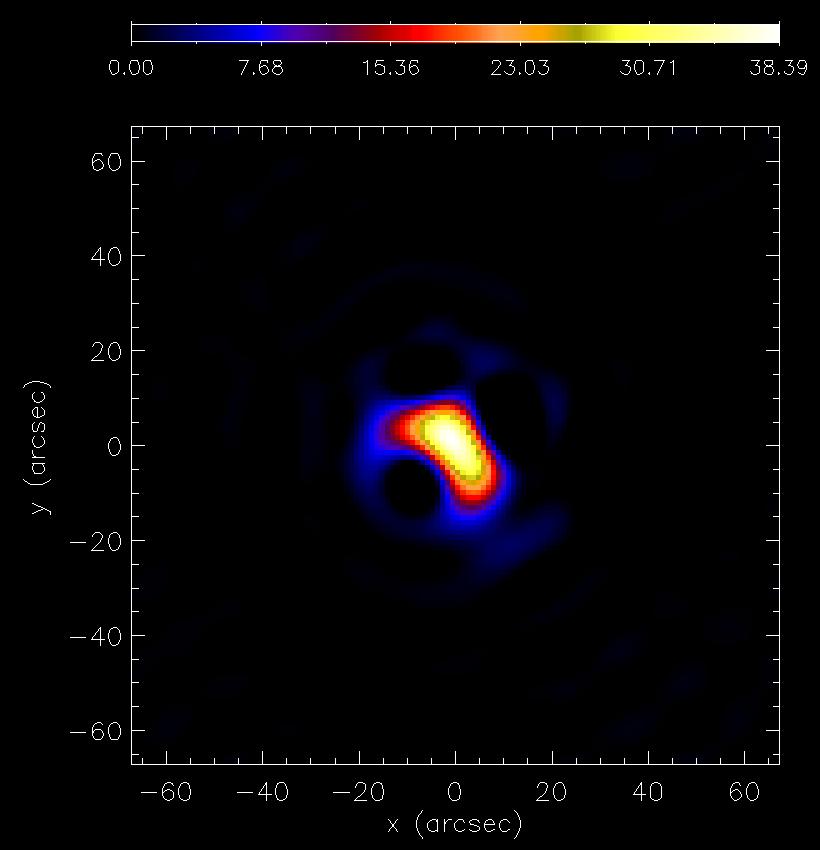}
  \includegraphics[scale=0.11]{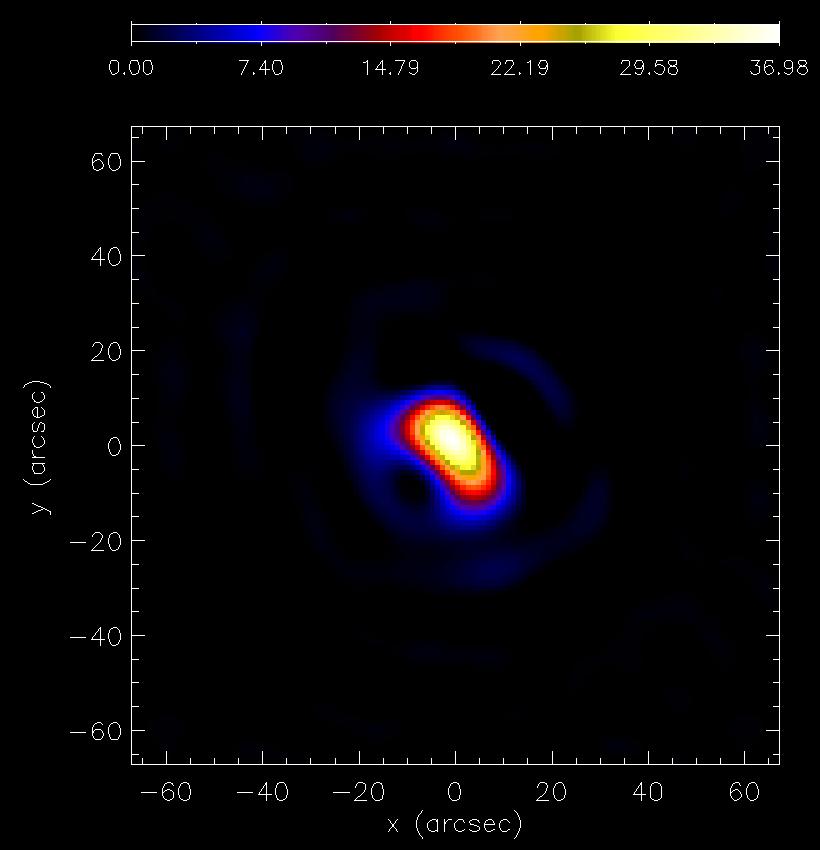}
  \includegraphics[scale=0.11]{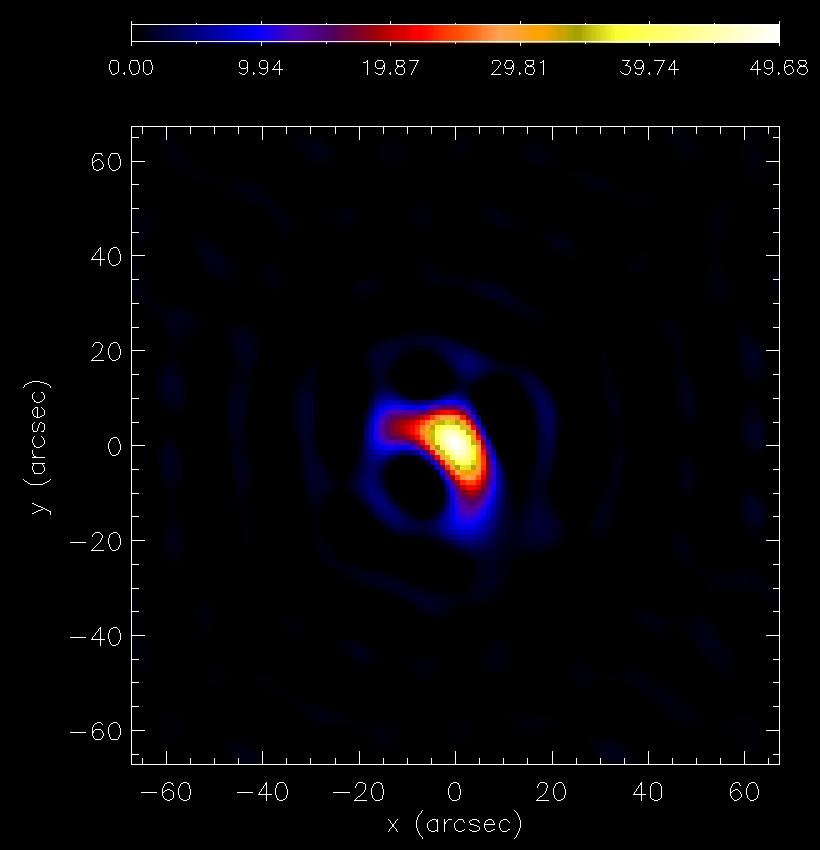}  \vskip 0.1cm   
  \includegraphics[scale=0.11]{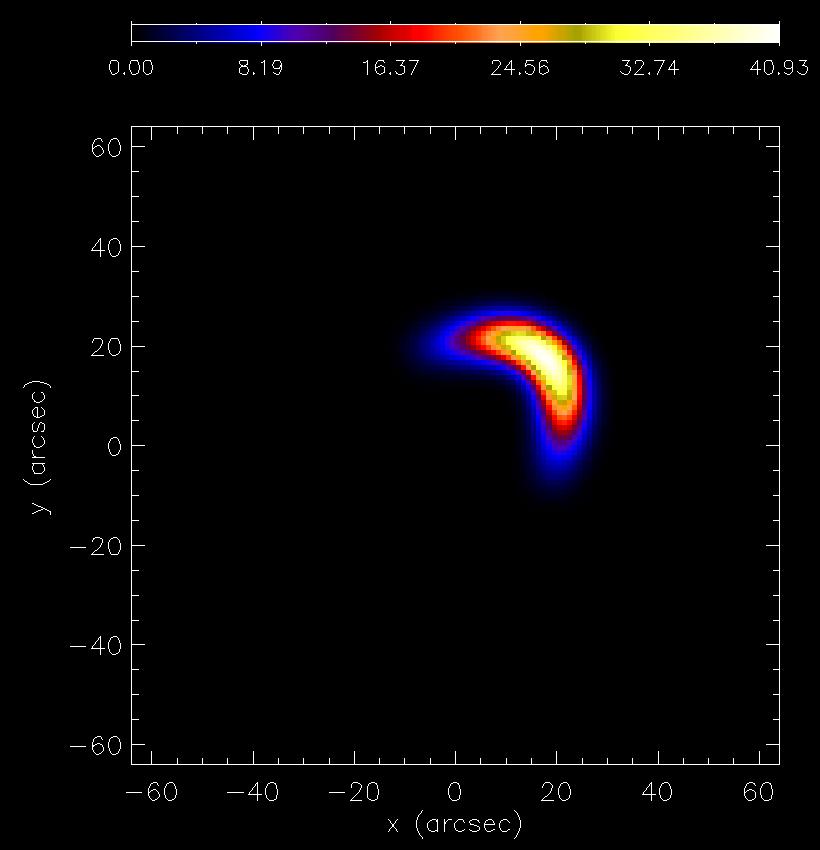}
  \includegraphics[scale=0.11]{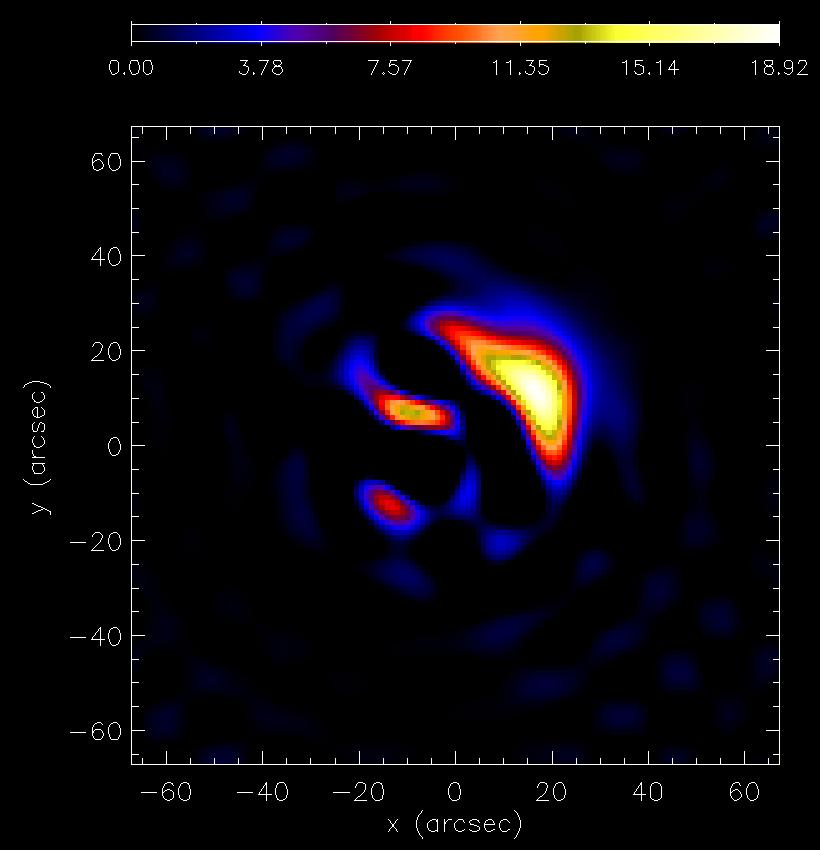}
  \includegraphics[scale=0.11]{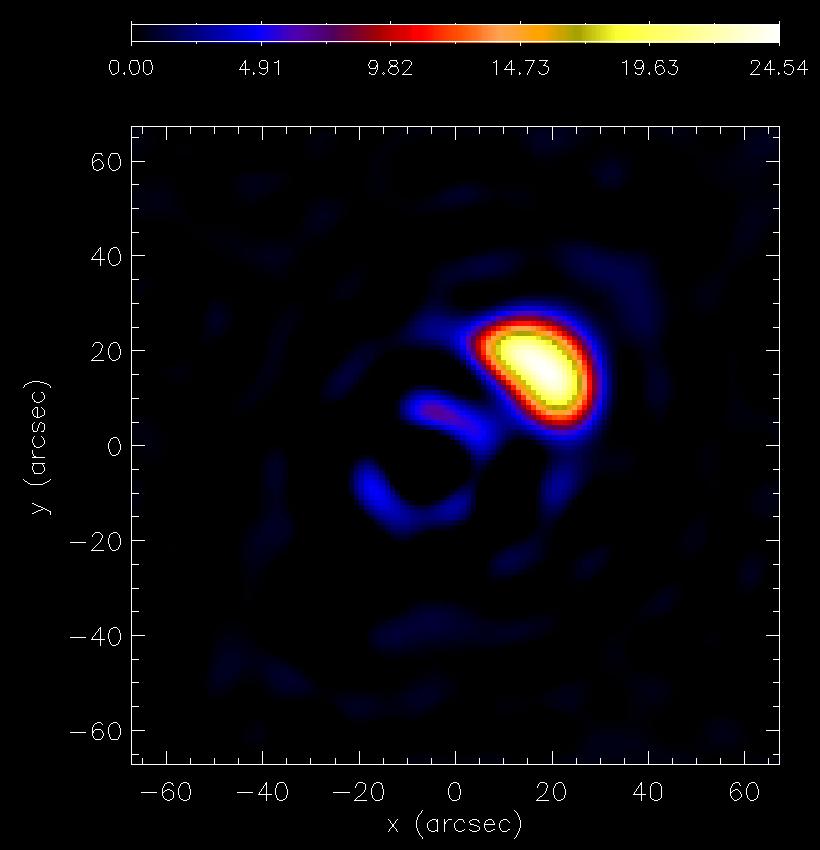}
  \includegraphics[scale=0.11]{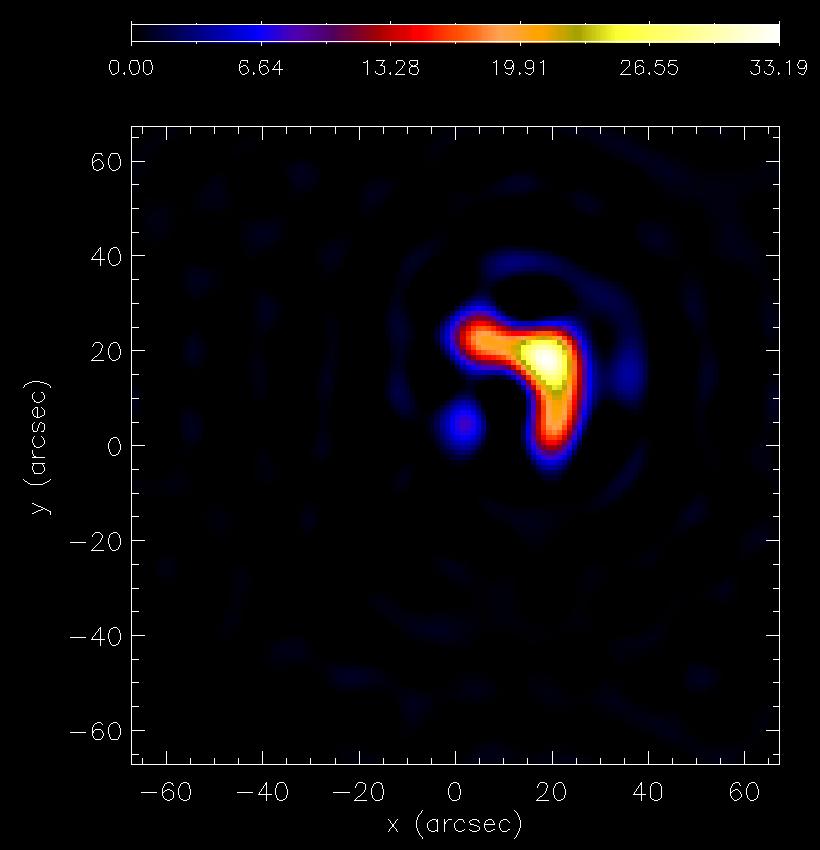}   
\caption{Reconstruction of four synthetic flaring configurations using simulated {\em{STIX}} visibilities. First column: ground-truth configurations. Second column: reconstructions provided by uv$\_$smooth. Third column: reconstructions obtained by using VSK-based interpolation when $\psi$ is the back projection (uv$\_$smooth$\_$BP). Third column: reconstructions obtained by using VSK-based interpolation when $\psi$ is the map of the CLEAN components (uv$\_$smooth$\_$CC). The ground-truth and reconstruction parameter values are in Tables 1--4.}
		\label{figmap} 
\end{figure}

We simulated four {\em{STIX}} configurations with an overall incident flux of $10^4$ photons cm$^{-2}$ s$^{-1}$ (see Figure \ref{figmap}, first column). The first two configurations (Configuration 1 and Configuration 2) consisted of two foot-points with centers located at two different positions along the main diagonal. The third and fourth configurations (Configuration 3 and Configuration 4) mimic two flaring loops, one at the center of the field-of-view and the other one off-center (refer to Tables \ref{tab1}--\ref{tab4} for details on the parameters of the four considered configurations).

Using the {\em{STIX}} simulation software we generated 25 realizations of synthetic {\em{STIX}} visibilities for each configuration. Then, Figure \ref{figmap} shows the results provided by the original version of uv$\_$smooth and by the two enhanced versions of the algorithm when the scaling functions are based upon the back-projected map (uv$\_$smooth$\_$BP) and the map of the CLEAN components (uv$\_$smooth$\_$CC). In Tables \ref{tab1}--\ref{tab4} the corresponding values of the reconstructed parameters are compared with the ones of the ground-truths, where for each parameter we have given the average value with respect to the 25 realizations and the corresponding standard deviation.

The CPU times employed to obtain the reconstructions are shown in Table \ref{cpu}. Tests have been carried out on a Intel(R) Core(TM) i7 CPU 4712MQ 2.13 GHz processor. 

\begin{table}[ht]
\caption{Results for the reconstruction of Configuration 1. The foot-point centers are denoted as $(x_p,y_p)$ while the flux is measured as photon cm$^{-2}$ s$^{-1}$.}
\begin{tabular}{lcccc}
	\hline 
    \hline
	& \multicolumn{4}{c}{First Peak} \\
        	\hline
	 &\hskip 0.1cm $x_p$ & \hskip 0.1cm $y_p$&   FWHM  & FLUX ($\times 10^3$) \\
	Simulated & -8.0 & -8.0 & 11.0 & 6.58 \\
	uv$\_$smooth  & -6.0  $\pm$ 0.6 & -5.0  $\pm$ 0.4 & 11.2 $\pm$ 0.3 & 5.08 $\pm$ 0.13\\
	uv$\_$smooth$\_$BP & -6.3 $\pm$ 0.4 & -6.2 $\pm$ 0.4 & 11.5 $\pm$ 0.4 & 5.53 $\pm$ 0.13 \\
    \smallskip
	uv$\_$smooth$\_$CC & -6.4  $\pm$ 0.5 & -6.0  $\pm$ 0.5 & 11.6  $\pm$ 0.4 & 5.52  $\pm$ 0.18 \\
	\hline 
	& \multicolumn{4}{c}{Second Peak}\\
        	\hline 
	 &\hskip 0.1cm $x_p$ & \hskip 0.1cm $y_p$&   FWHM  & FLUX ($\times 10^3$) \\ 
	Simulated & 8.0 & 8.0 & 11.0 & 3.21 \\
	uv$\_$smooth  & 8.0  $\pm$ 0.5 & 6.4  $\pm$ 0.5 & 10.7 $\pm$ 0.5 & 2.42 $\pm$ 0.12\\
	uv$\_$smooth$\_$BP & 8.1  $\pm$ 0.4 & 8.3  $\pm$ 0.6 & 11.5  $\pm$ 0.5 & 2.70  $\pm$ 0.13 \\
	\smallskip
	uv$\_$smooth$\_$$\_$CC    & 7.9  $\pm$ 0.6 & 6.9  $\pm$ 0.3 & 12.3 $\pm$ 0.8 & 2.77 $\pm$ 0.14 \\
\end{tabular}
\begin{tabular}{lc}
	\hline 
	&  Total Flux ($\times 10^3$) \\
	\hline 
	Simulated & 10.00   \\
   uv$\_$smooth & 9,27 $\pm$ 0.18\\
	uv$\_$smooth$\_$BP  & 9.86 $\pm$ 0.23\\
		\smallskip
	uv$\_$smooth$\_$CC & 10.10 $\pm$ 0.19\\
	\hline
	 \hline
\end{tabular}
\label{tab1}
\end{table}

\begin{table}[ht]
\caption{Results for the reconstruction of Configuration 2. The foot-point centers are denoted as $(x_p,y_p)$ while the flux is measured as photon cm$^{-2}$ s$^{-1}$.}
\begin{tabular}{lcccc}
	\hline 
	\hline
	& \multicolumn{4}{c}{First Peak} \\
        	\hline 
	 &\hskip 0.2cm $x_p$ & \hskip 0.2cm $y_p$&   FWHM  & FLUX ($\times 10^3$) \\ 
	Simulated & -24.0 & -24.0 & 11.0 & 6.51 \\
 	uv$\_$smooth  & -7.5 $\pm$ 0.5 & -10.6 $\pm$ 0.4 & 12.3 $\pm$ 0.1 & 8.23 $\pm$ 0.19 \\
 	uv$\_$smooth$\_$BP & -21.9 $\pm$ 0.2 & -21.7 $\pm$ 0.4 & 10.8 $\pm$ 0.2 & 5.26 $\pm$ 0.11 \\
 		\smallskip
 	uv$\_$smooth$\_$CC & -21.8 $\pm$ 0.3 & -21.7 $\pm$ 0.4 & 11.4 $\pm$ 0.4 & 5.27 $\pm$ 0.12 \\
	\hline 
	& \multicolumn{4}{c}{Second Peak}\\
    \hline 
	&\hskip 0.2cm $x_p$ & \hskip 0.2cm $y_p$&   FWHM  & FLUX ($\times 10^3$) \\ 
	Simulated & 24.0 & 24.0 & 11.0 & 3.25 \\
	uv$\_$smooth &  8.5  $\pm$ 0.5 & -9.4 $\pm$ 0.8 & 12.9 $\pm$ 0.1 & 8.94 $\pm$ 0.26 \\	
	uv$\_$smooth$\_$BP& 24.0 $\pm$ 0.2 & 24.4 $\pm$ 0.5 & 10.9 $\pm$ 0.2 & 2.50 $\pm$ 0.12 \\
		\smallskip
	uv$\_$smooth$\_$CC & 23.3 $\pm$ 0.3 & 24.6 $\pm$ 0.8 & 10.9 $\pm$ 0.7 & 2.35 $\pm$ 0.14 \\
\end{tabular}
\begin{tabular}{lc}
	\hline 
	&  Total Flux ($\times 10^3$) \\
	\hline \\
	Simulated & 10.00    \\
    uv$\_$smooth & 9.88 $\pm$ 0.34  \\	
    uv$\_$smooth$\_$BP & 10.55 $\pm$ 0.30  \\
		\smallskip
	uv$\_$smooth$\_$CC & 12.37 $\pm$ 0.28  \\
	\hline
	\hline
\end{tabular}
\label{tab2}
\end{table}

\begin{table}[ht]
\caption{Results for the reconstruction of Configuration 3. The position of the pixel with maximum intensity is denoted as $(x_p,y_p)$. The flux units are photon cm$^{-2}$ s$^{-1}$.}
\begin{tabular}{lccc}
	\hline
	\hline
	 &$x_p$ & $y_p$ &  Total Flux ($\times 10^3$) \\
	\hline 	 
	Simulated &  0.0 & 0.0 & 10.00   \\
	uv$\_$smooth &  0.7 $\pm$ 0.6 & 1.3  $\pm$ 0.5 & 9.71  $\pm$ 0.27 \\	
	uv$\_$smooth$\_$BP & -0.5 $\pm$ 0.6 & 1.4 $\pm$ 0.6 & 10.55 $\pm$ 0.02\\
		\smallskip   
	uv$\_$smooth$\_$CC & -0.6 $\pm$ 0.5 & 0.6 $\pm$ 0.4 & 10.55 $\pm$ 0.02\\
	\hline
	\hline
\end{tabular}
\label{tab3}
\end{table}

\begin{table}[ht]
\caption{Results for the reconstruction of Configuration 4. The position of the pixel with maximum intensity is denoted as $(x_p,y_p)$. The flux units are photon cm$^{-2}$ s$^{-1}$.}
\begin{tabular}{lccc}
	\hline 
    \hline
	 &$x_p$ & $y_p$&  Total Flux ($\times 10^3$) \\
	\hline  
	Simulated & 18.0 & 18.0 & 10.00   \\
	uv$\_$smooth & 15.4  $\pm$ 0.8 & 11.4  $\pm$ 0.7 & 10.59  $\pm$ 0.01 \\	
	uv$\_$smooth$\_$BP & 16.6 $\pm$ 0.7 & 15.8 $\pm$ 0.8 & 10.59 $\pm$ 0.02\\
		\smallskip   
	uv$\_$smooth$\_$CC & 17.0 $\pm$ 0.7 & 16.8 $\pm$ 0.8 & 10.59 $\pm$ 0.01\\
	\hline
	\hline
\end{tabular}
\label{tab4}
\end{table}

\begin{table}[ht]
\caption{CPU burden (in second) employed by the three reconstruction algorithms averaged over the data corresponding to the four configurations.}
\begin{tabular}{lc}
	\hline
	\hline
	 & CPU times \\
	\hline 
    uv$\_$smooth & 0.18 \\  
	uv$\_$smooth$\_$BP & 4.24 \\
		\smallskip  
	uv$\_$smooth$\_$CC & 6.78 \\
	\hline
	\hline
\end{tabular}
    \label{cpu}
\end{table}

\subsection{RHESSI observations}

\begin{figure}[h!]
\centering
     \includegraphics[scale=0.11]{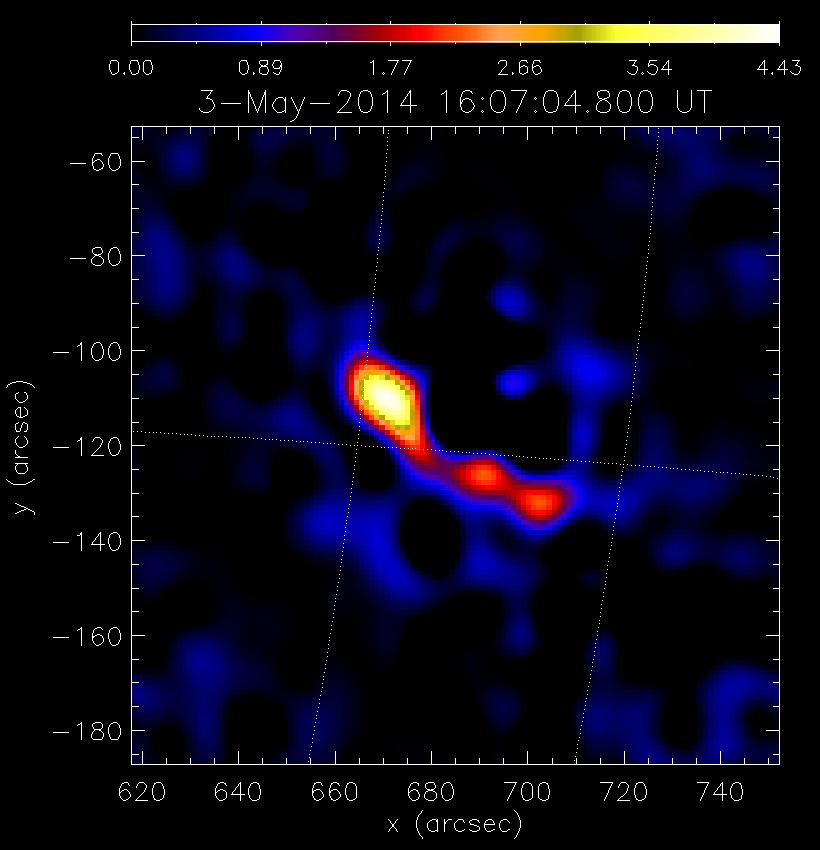}   
    \includegraphics[scale=0.11]{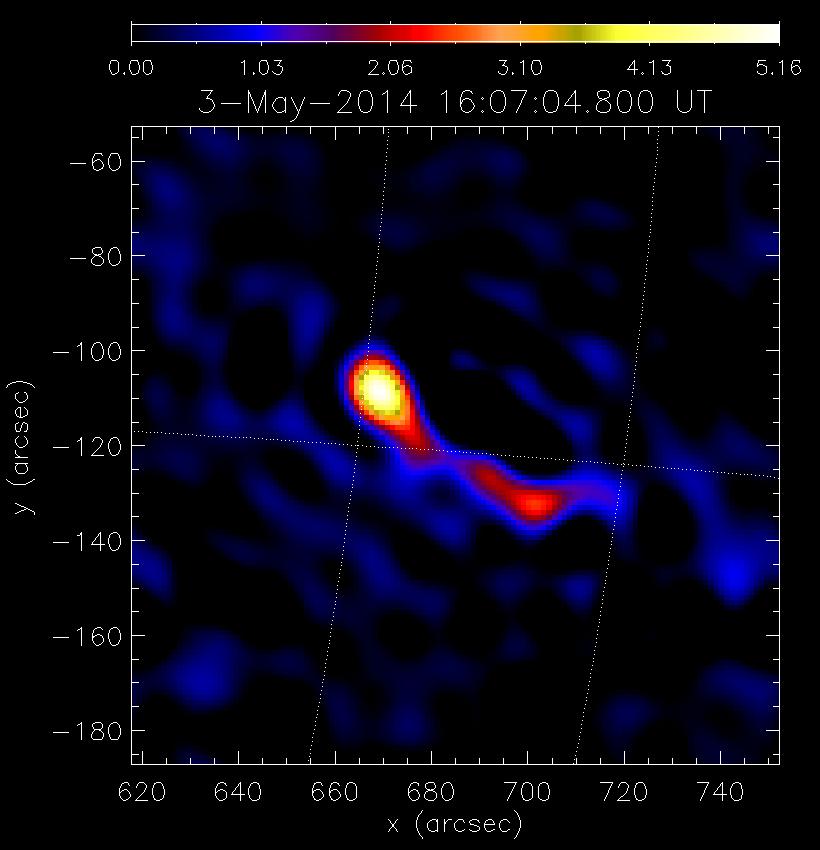}
    \includegraphics[scale=0.11]{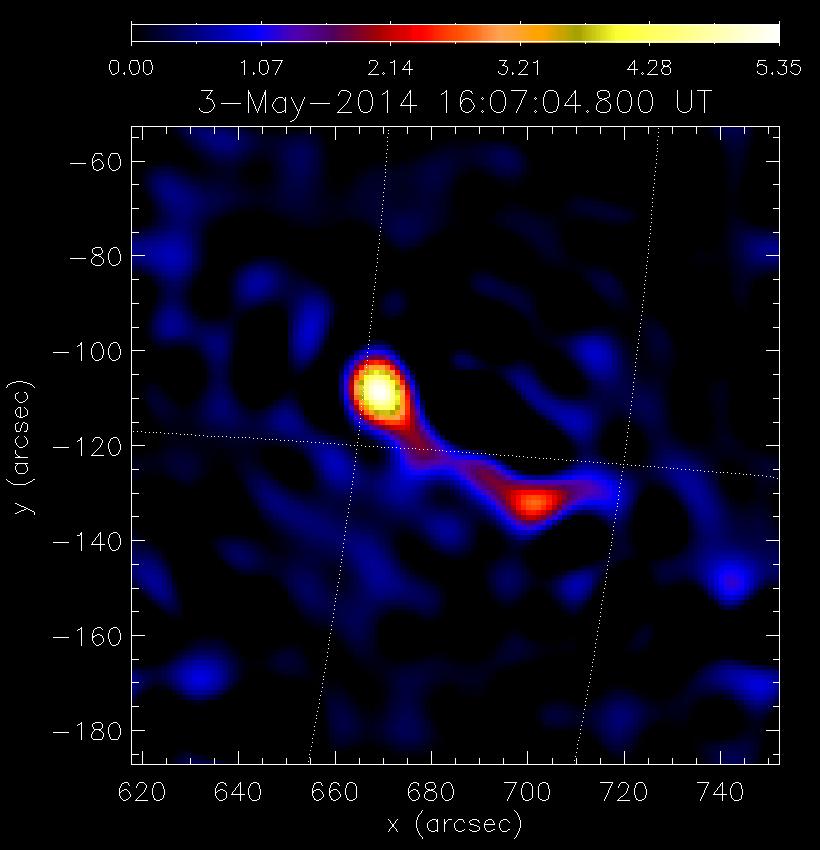}
    \includegraphics[scale=0.11]{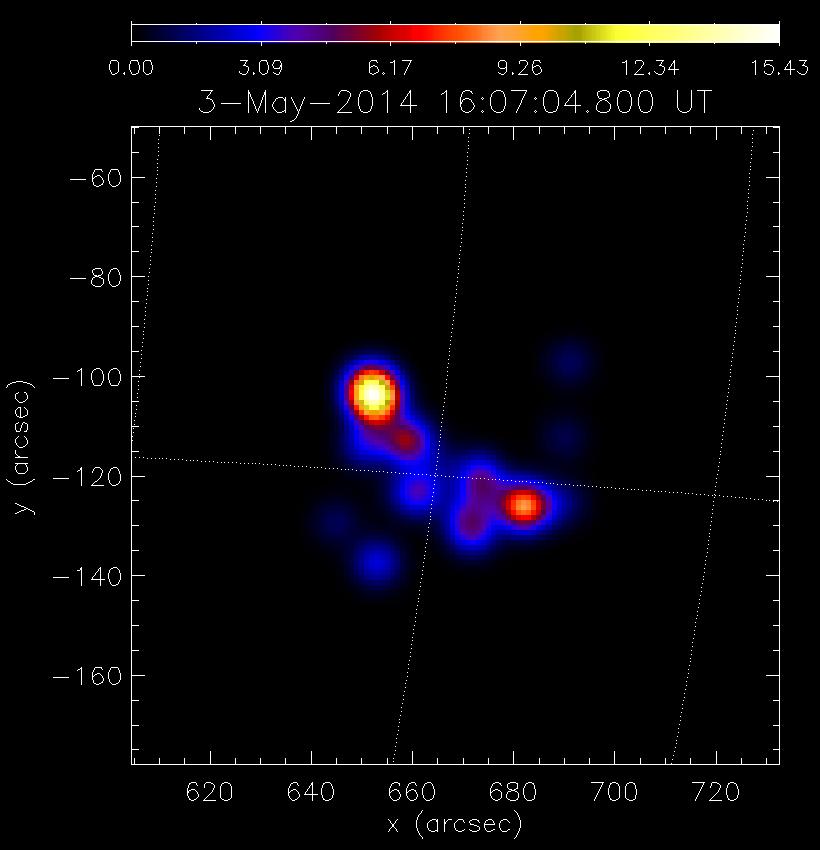} \vskip 0.1cm  
     \includegraphics[scale=0.11]{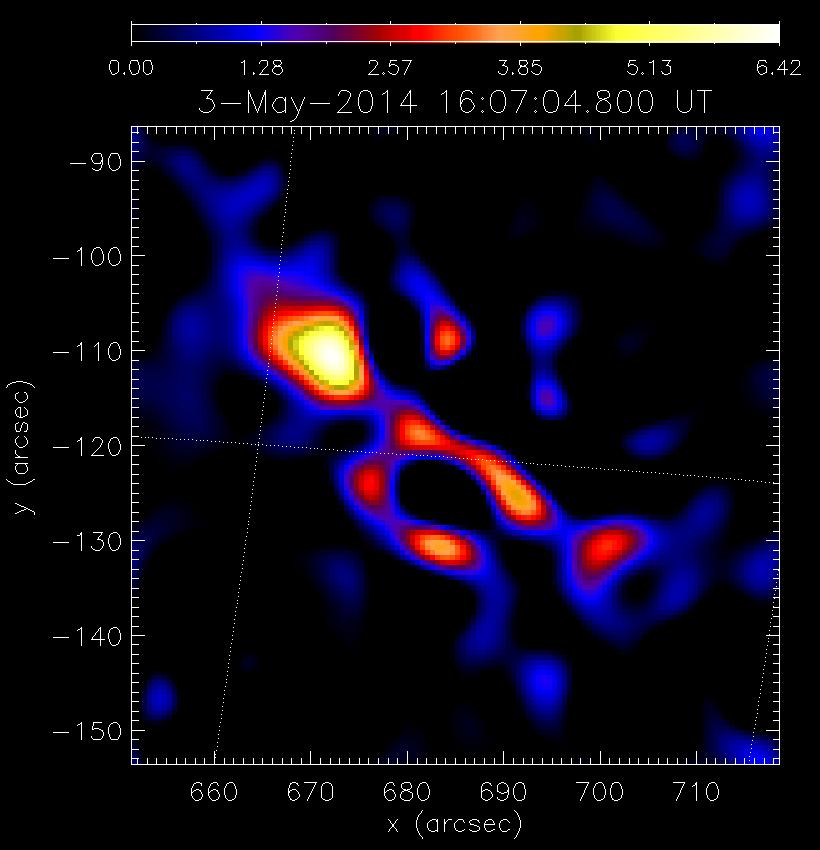}   
    \includegraphics[scale=0.11]{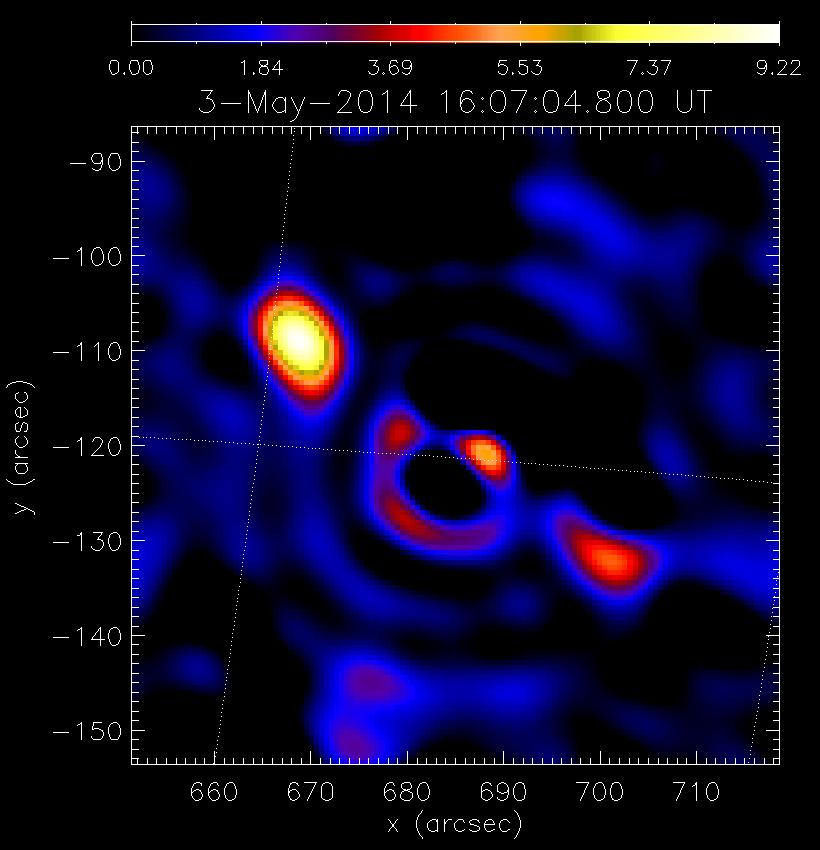}
    \includegraphics[scale=0.11]{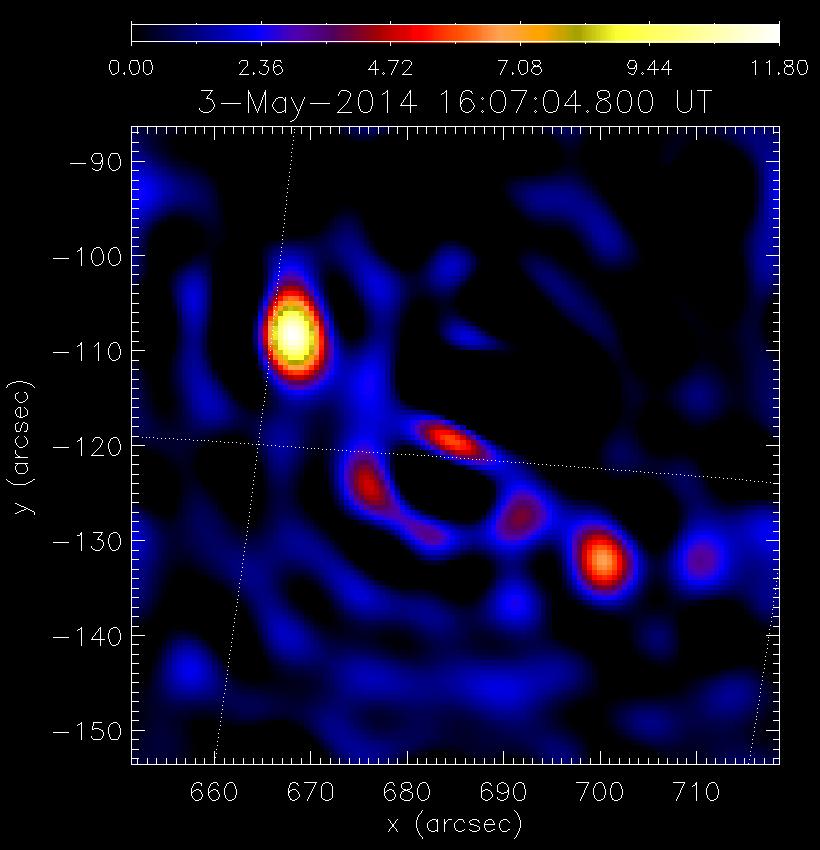}
    \includegraphics[scale=0.11]{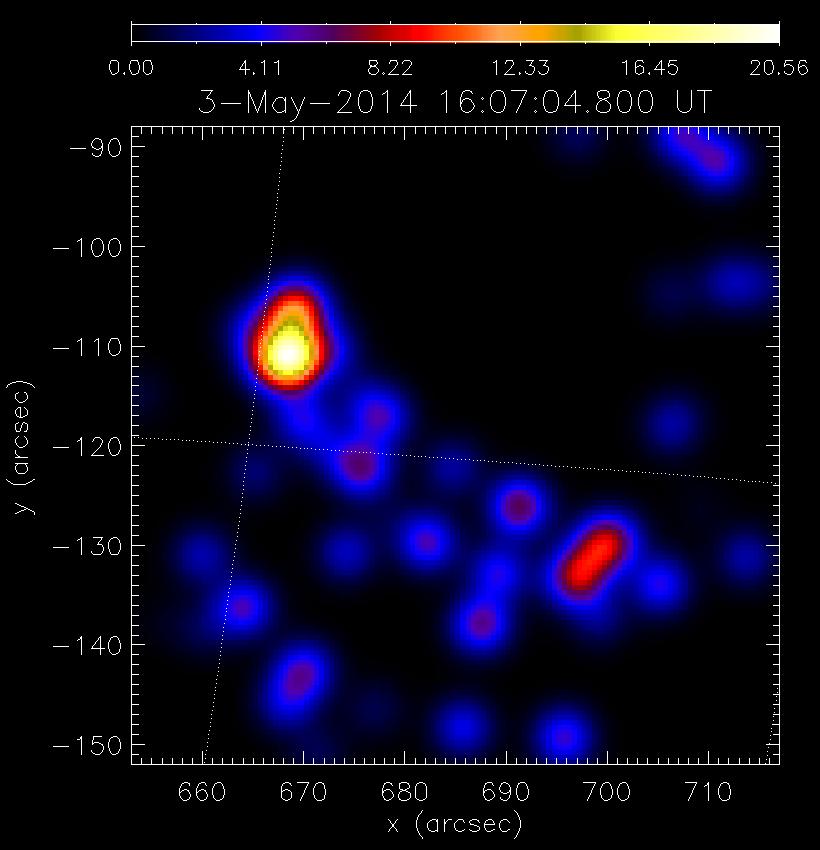} \vskip 0.1cm  
     \includegraphics[scale=0.11]{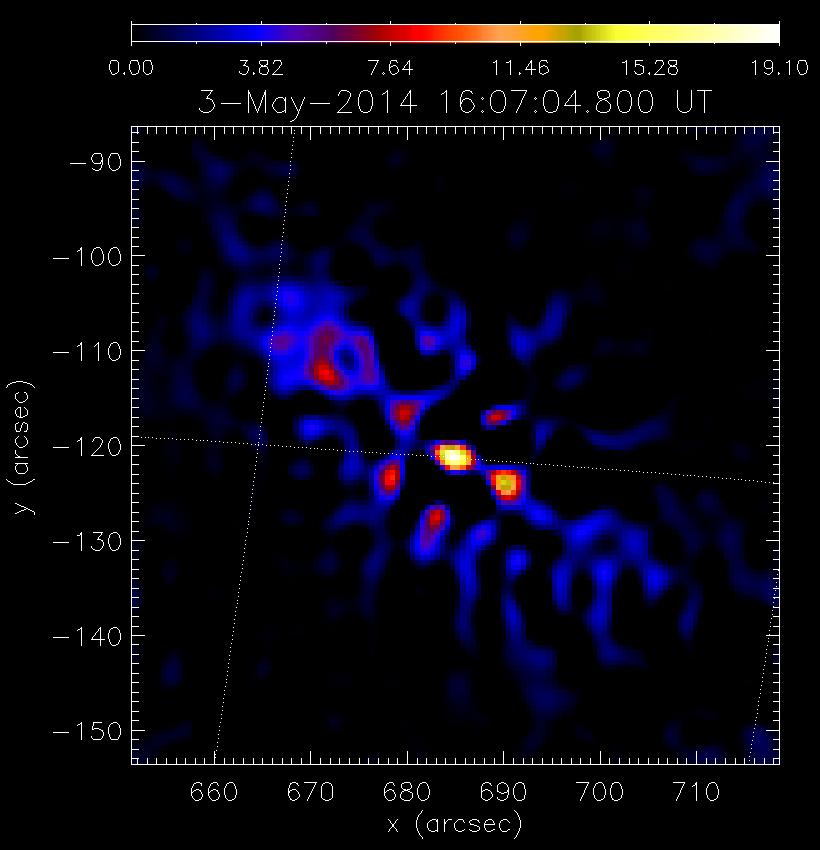}   
    \includegraphics[scale=0.11]{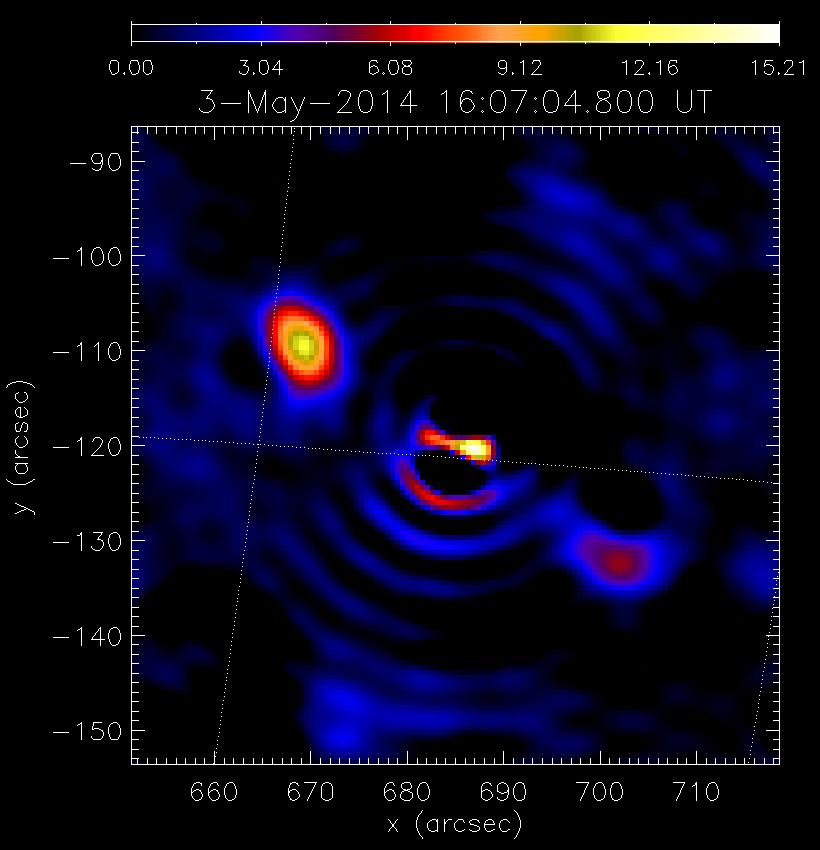}
    \includegraphics[scale=0.11]{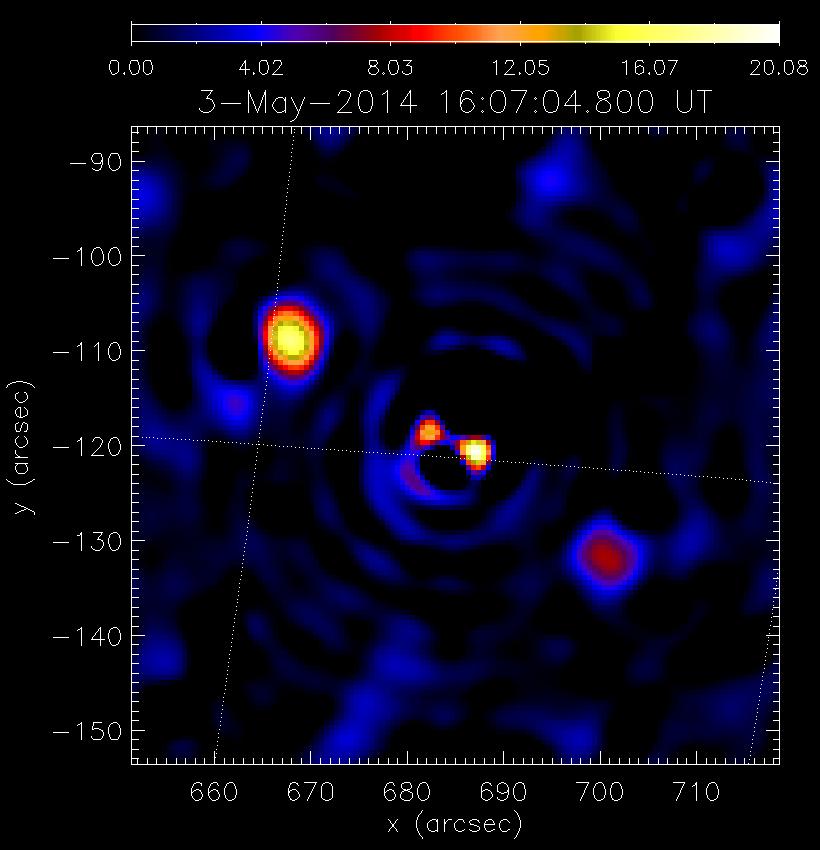}
    \includegraphics[scale=0.11]{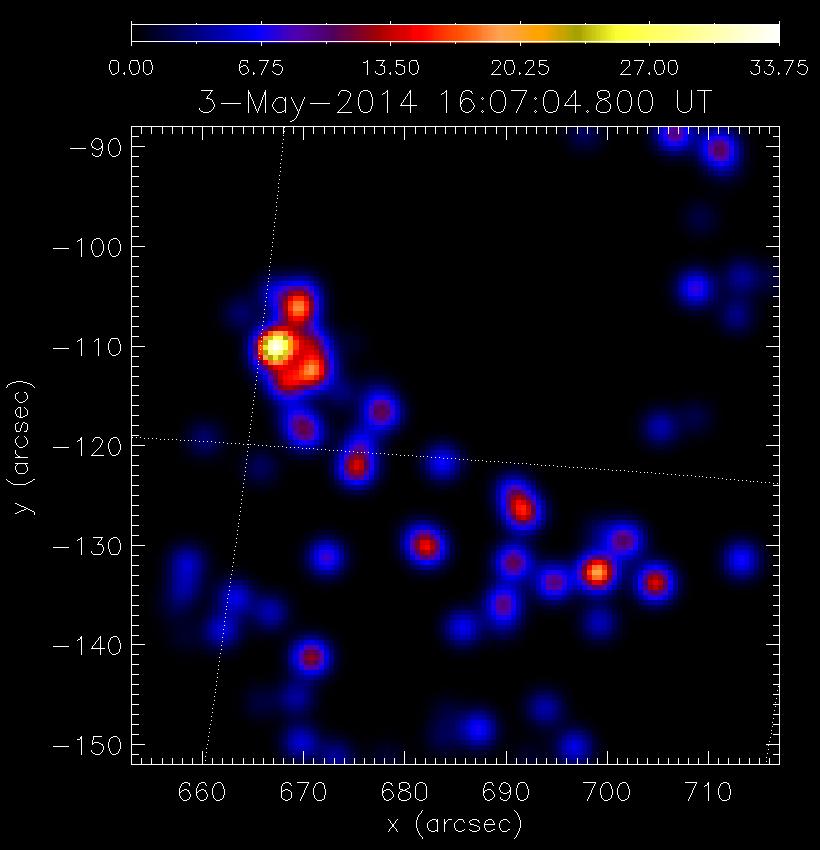}
    \caption{Reconstruction of the flare observed by RHESSI on May 3 in 2014 at 16:07:04 UT. From left to right, the columns contain the reconstructions via uv$\_$smooth, uv$\_$smooth$\_$BP, uv$\_$smooth$\_$CC and CLEAN. From top to bottom the three rows indicate the reconstructions obtained using {\em{RHESSI}} detectors 3 through 9, 2 through 9 and 1 through 9, respectively.
}
    \label{fig_rhessi_maydet}
\end{figure}

\begin{figure}[h!]
\centering
     \includegraphics[scale=0.128]{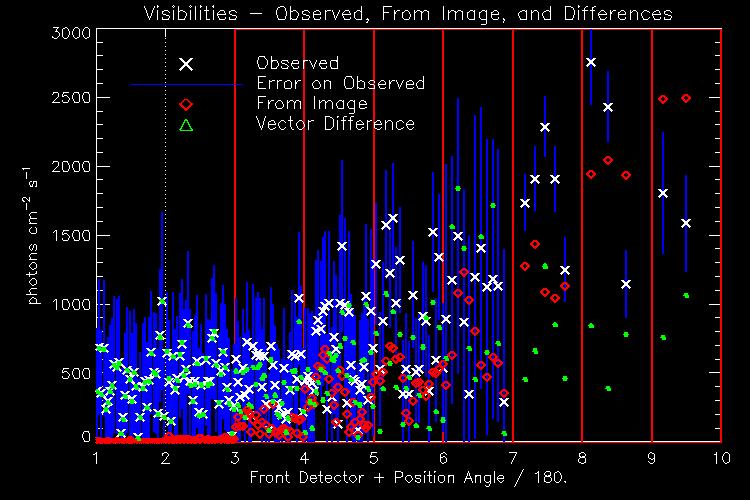}   
    \includegraphics[scale=0.128]{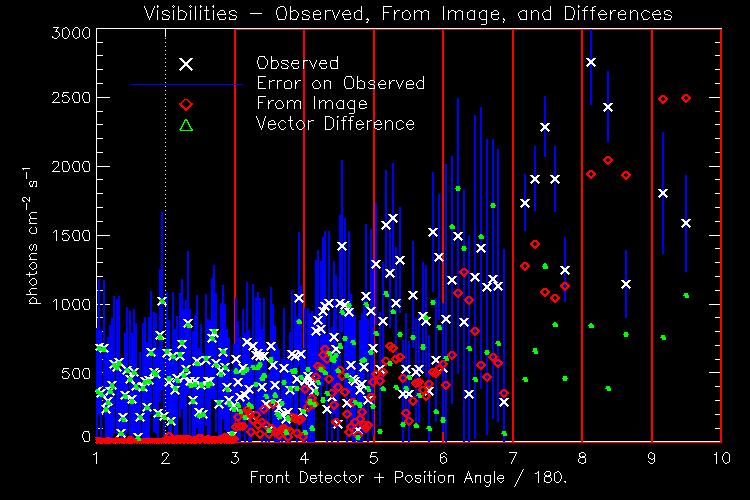}
    \includegraphics[scale=0.128]{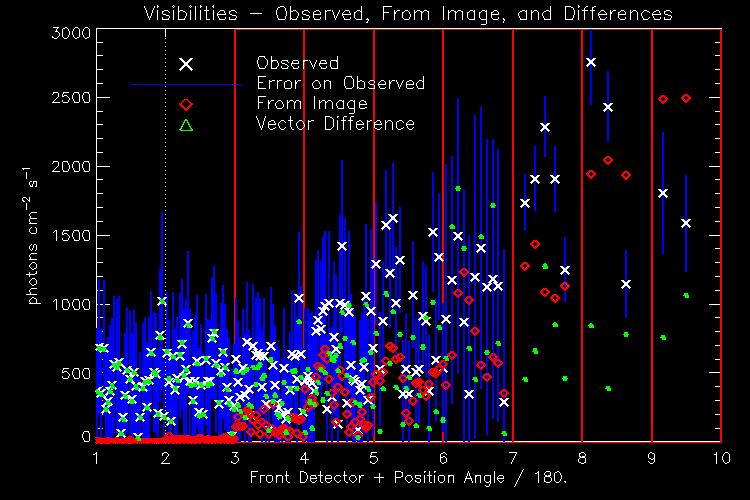}
    \includegraphics[scale=0.128]{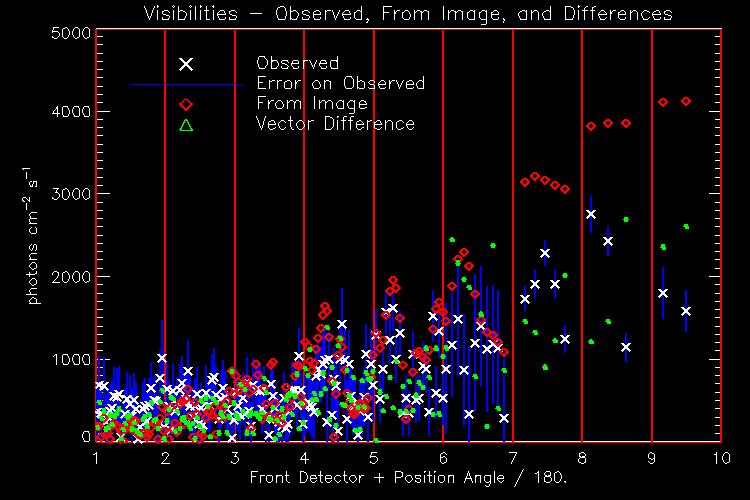} \vskip 0.1cm  
     \includegraphics[scale=0.128]{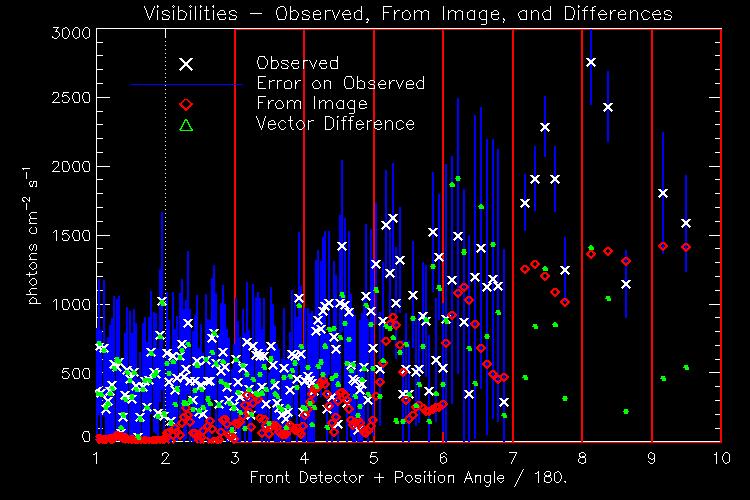}   
    \includegraphics[scale=0.128]{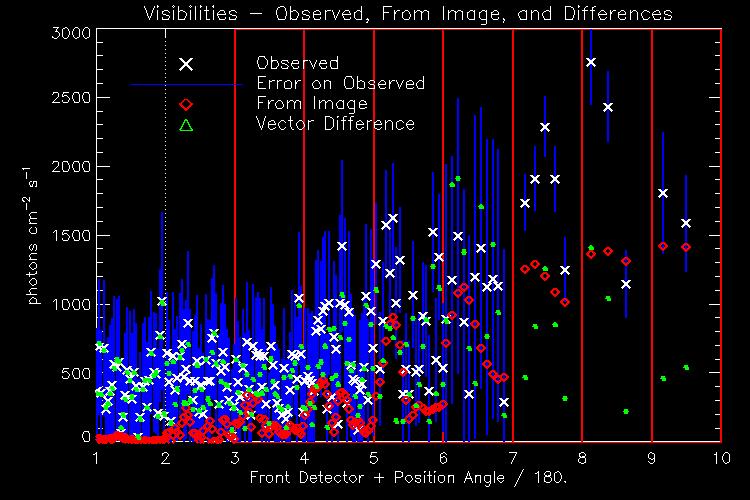}
    \includegraphics[scale=0.128]{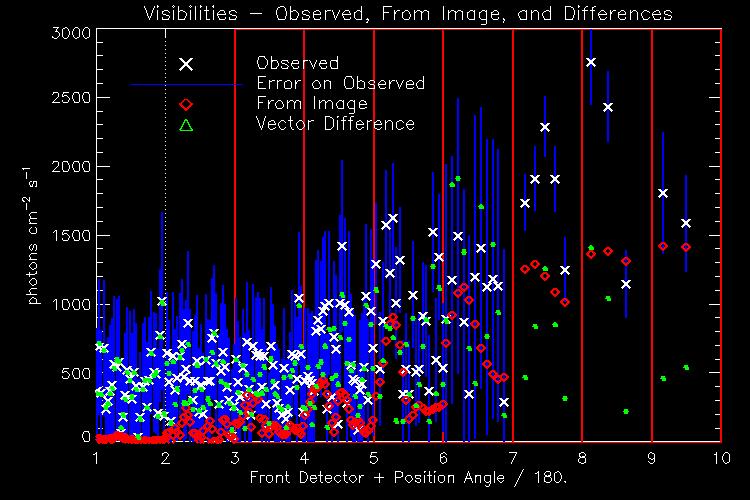}
    \includegraphics[scale=0.128]{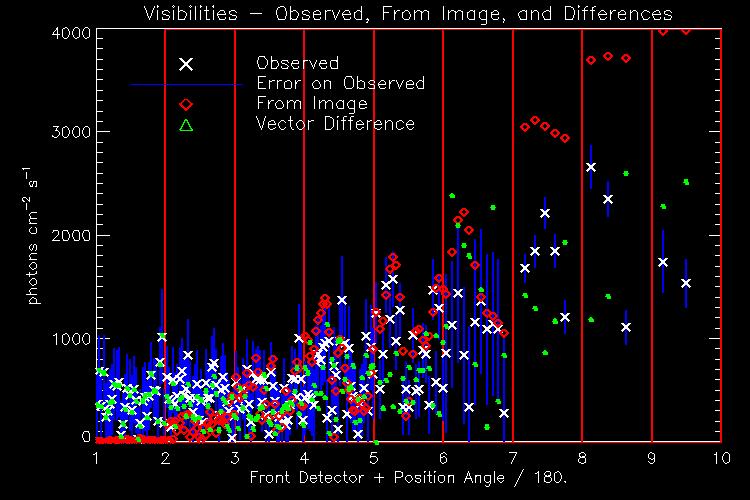} \vskip 0.1cm  
     \includegraphics[scale=0.128]{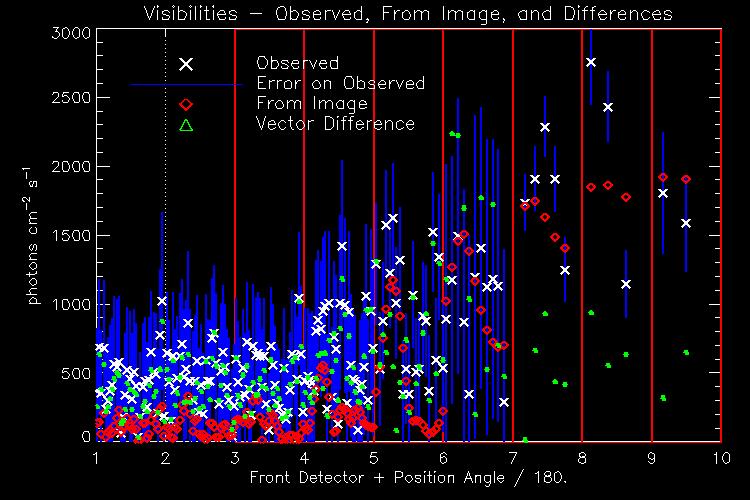}   
    \includegraphics[scale=0.128]{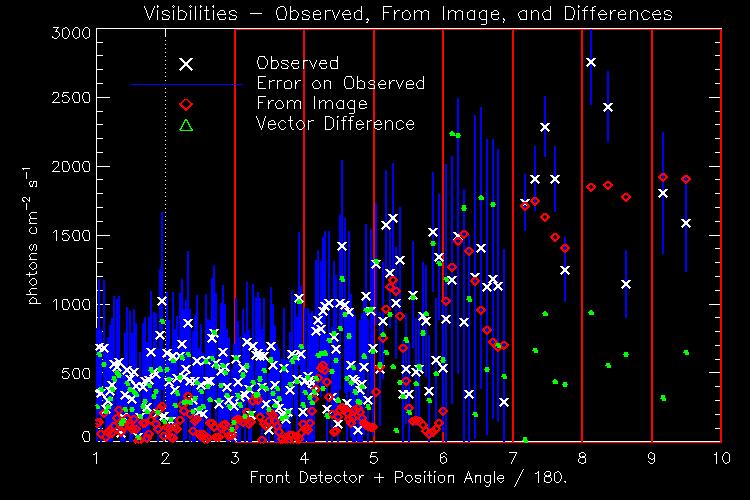}
    \includegraphics[scale=0.128]{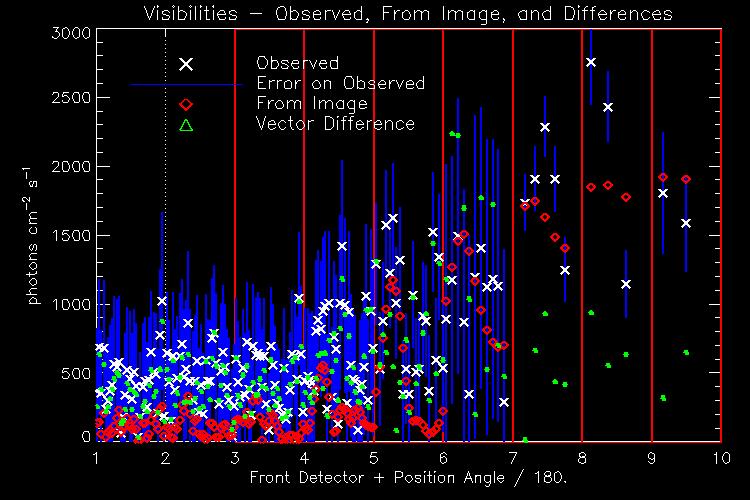}
    \includegraphics[scale=0.128]{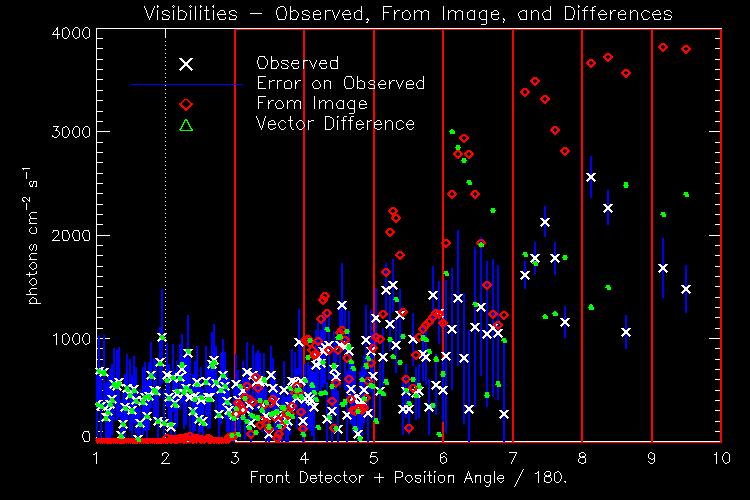}
    \caption{Comparison between predicted and measured visibilities for the flare observed by RHESSI on May 3 2014 at 16:07:04 UT. From left to right, the columns contain the fits corresponding to uv$\_$smooth, uv$\_$smooth$\_$BP, uv$\_$smooth$\_$CC and CLEAN. From top to bottom, the rows correspond to using detector configurations from 3 through 9, from 2 through 9, and from 1 through 9, respectively.
}
    \label{fig_rhessi_vis_dets}
\end{figure}

\begin{figure}[h!]
\centering
   \includegraphics[scale=0.11]{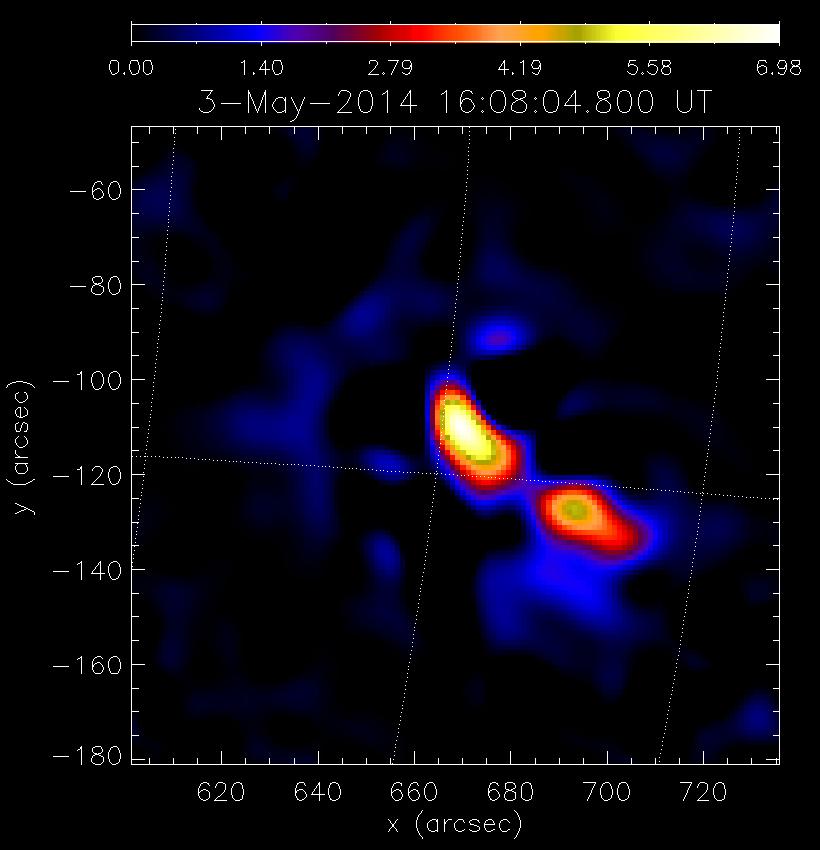}   
    \includegraphics[scale=0.11]{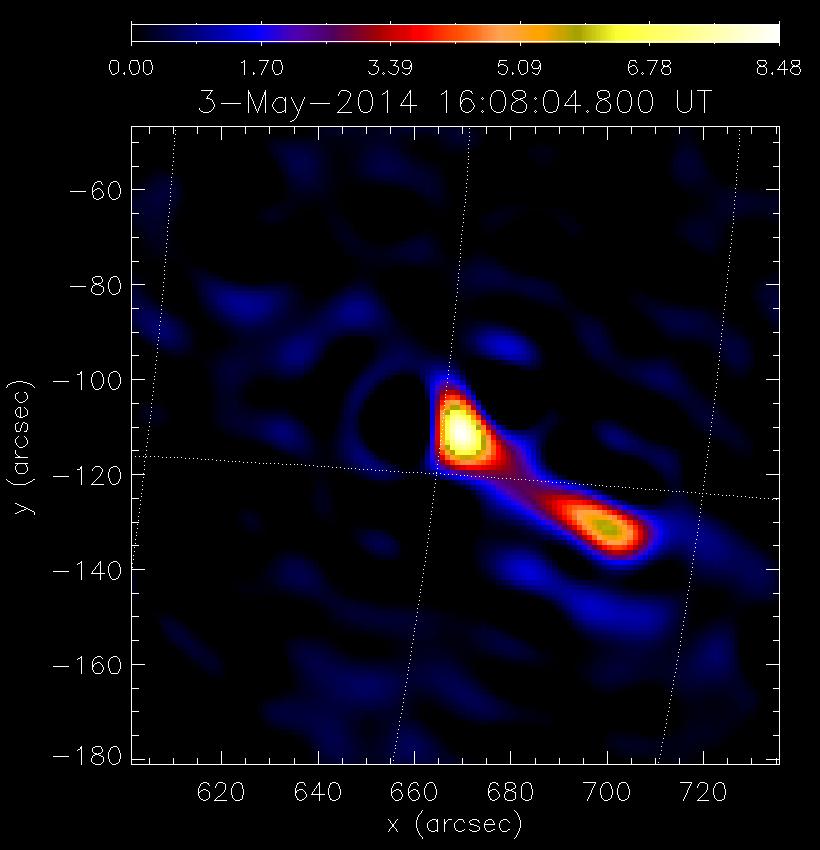} 
    \includegraphics[scale=0.11]{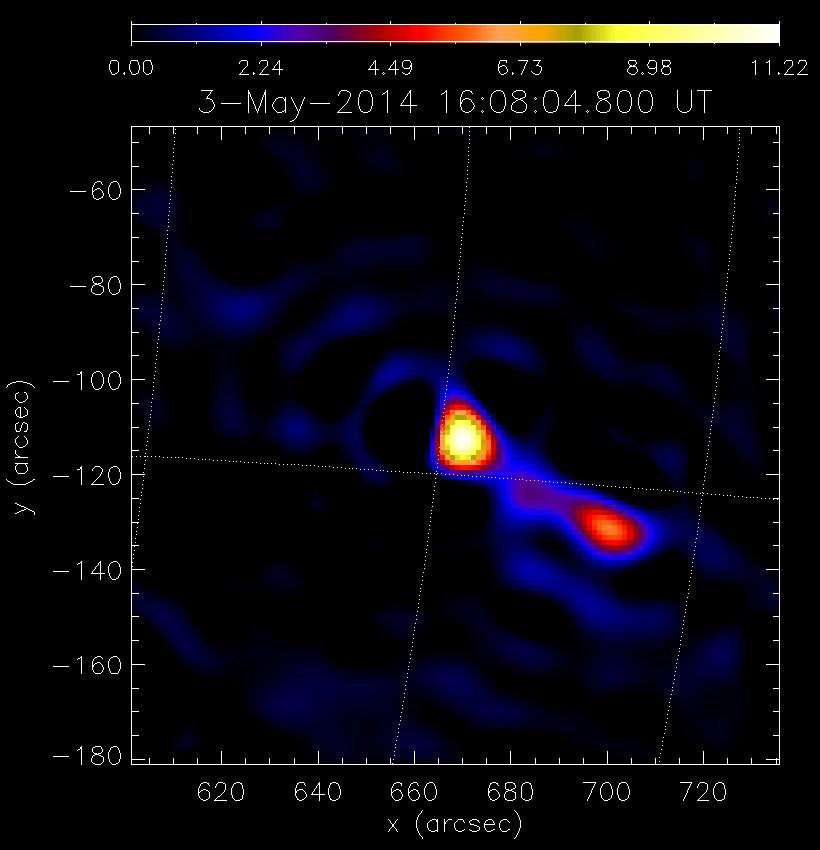} 
    \includegraphics[scale=0.11]{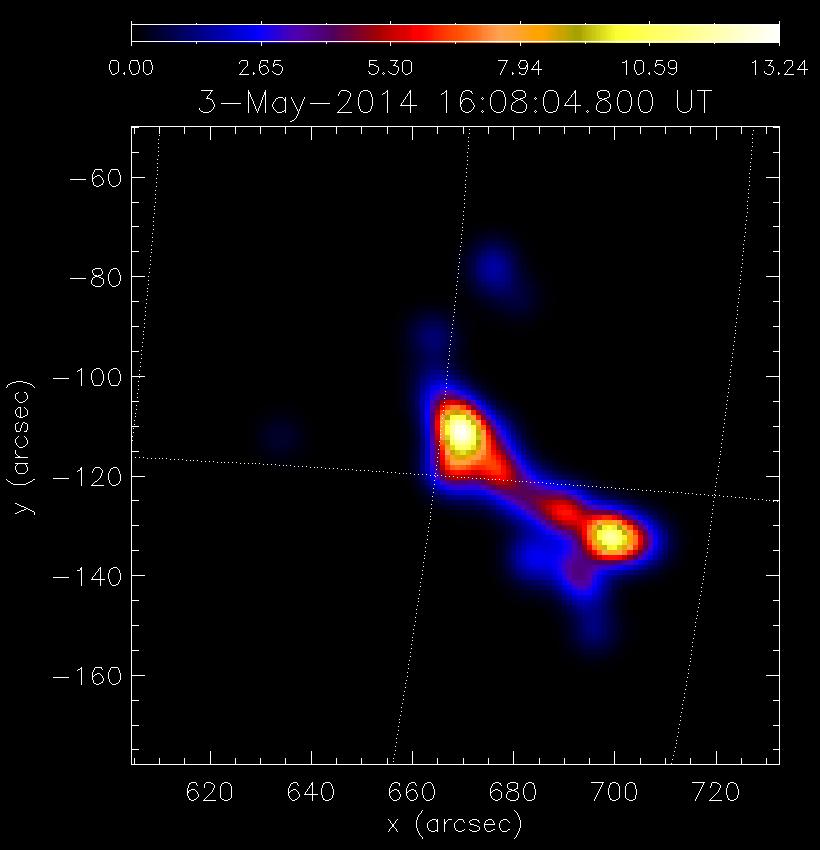} \vskip 0.1cm       
     \includegraphics[scale=0.11]{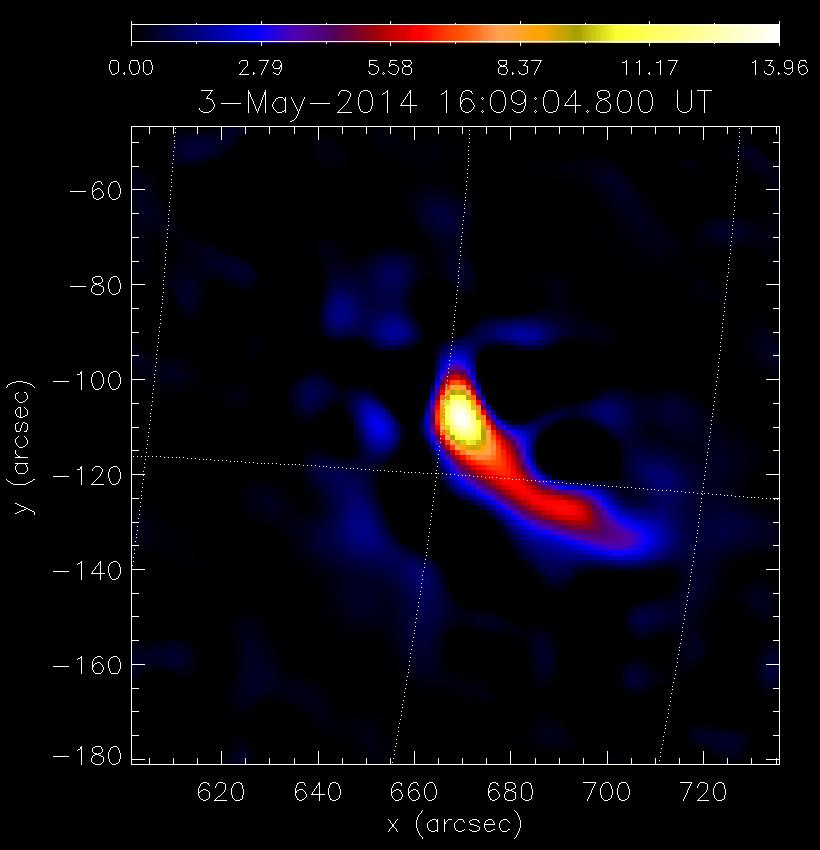}   
    \includegraphics[scale=0.11]{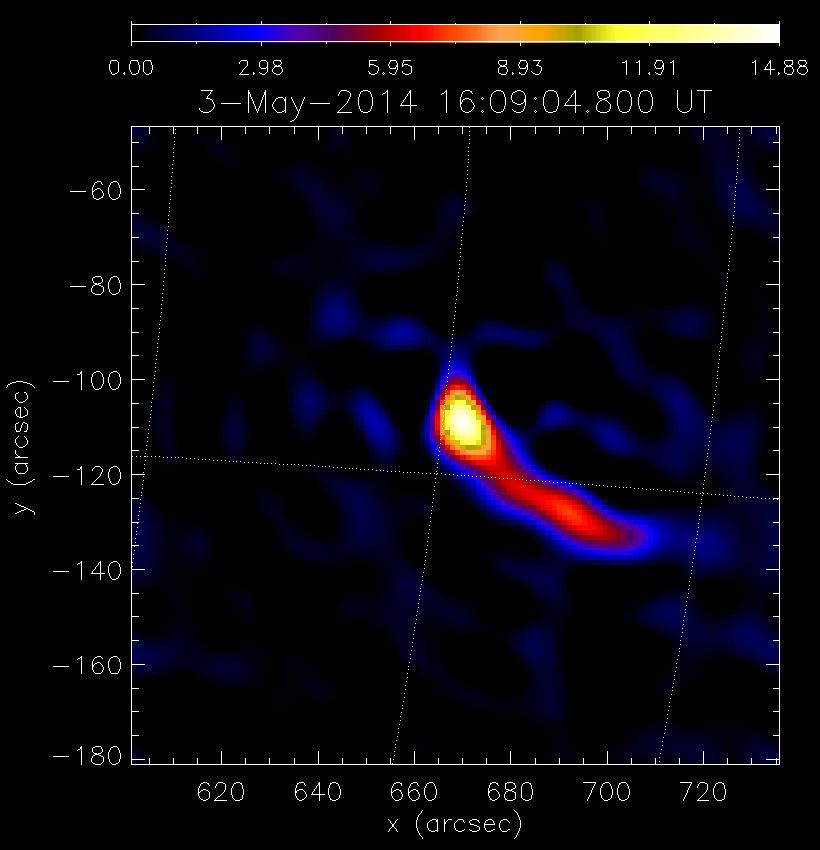} 
    \includegraphics[scale=0.11]{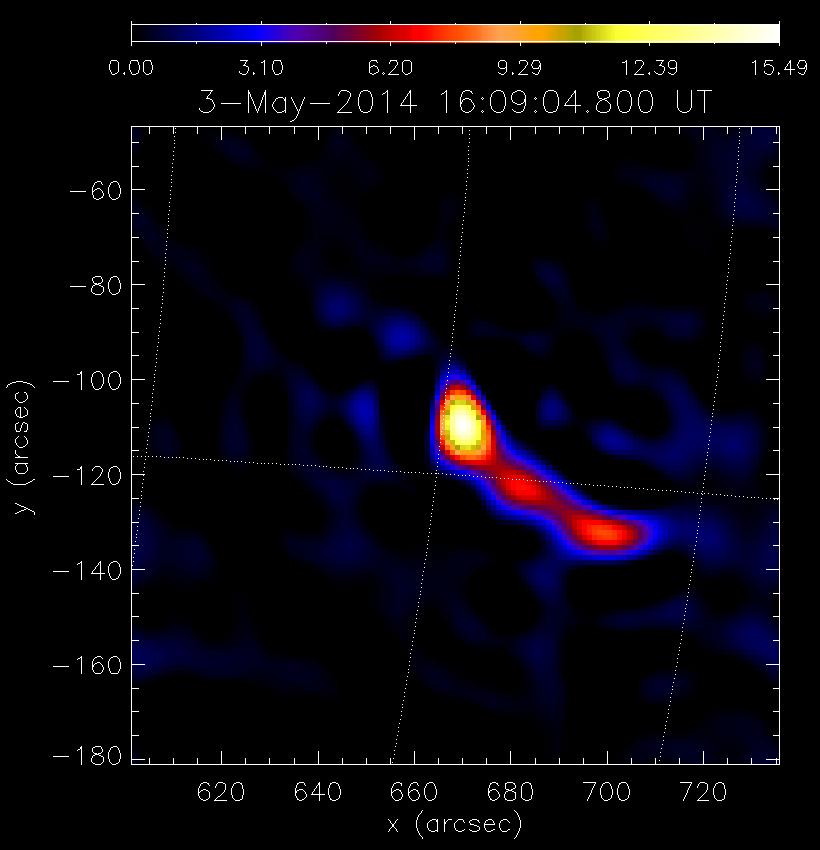} 
    \includegraphics[scale=0.11]{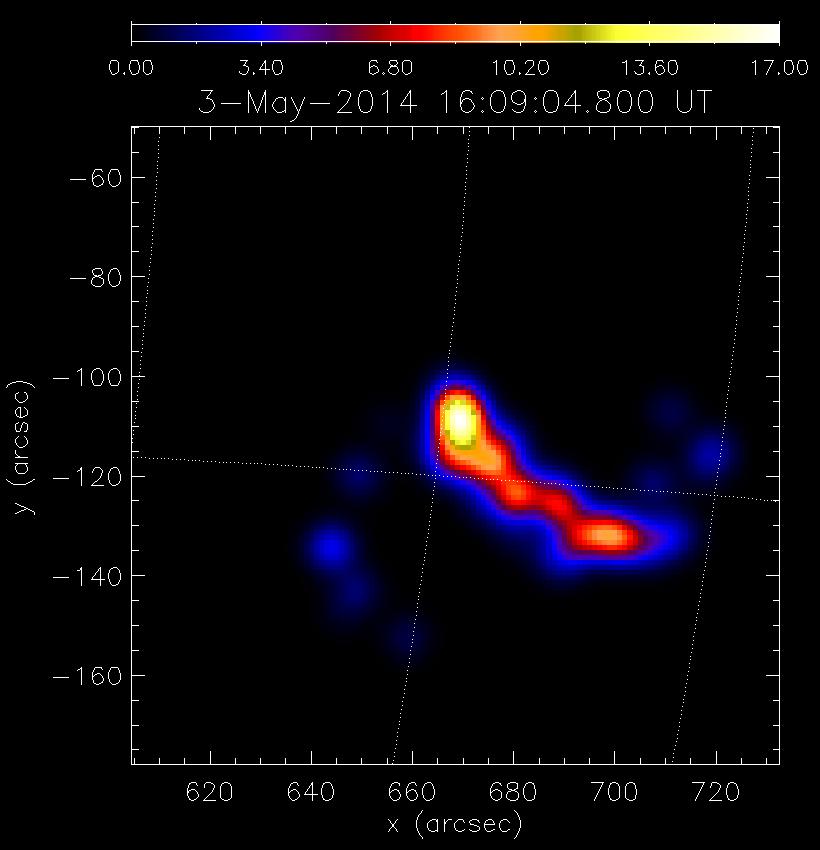} \vskip 0.1cm   
     \includegraphics[scale=0.11]{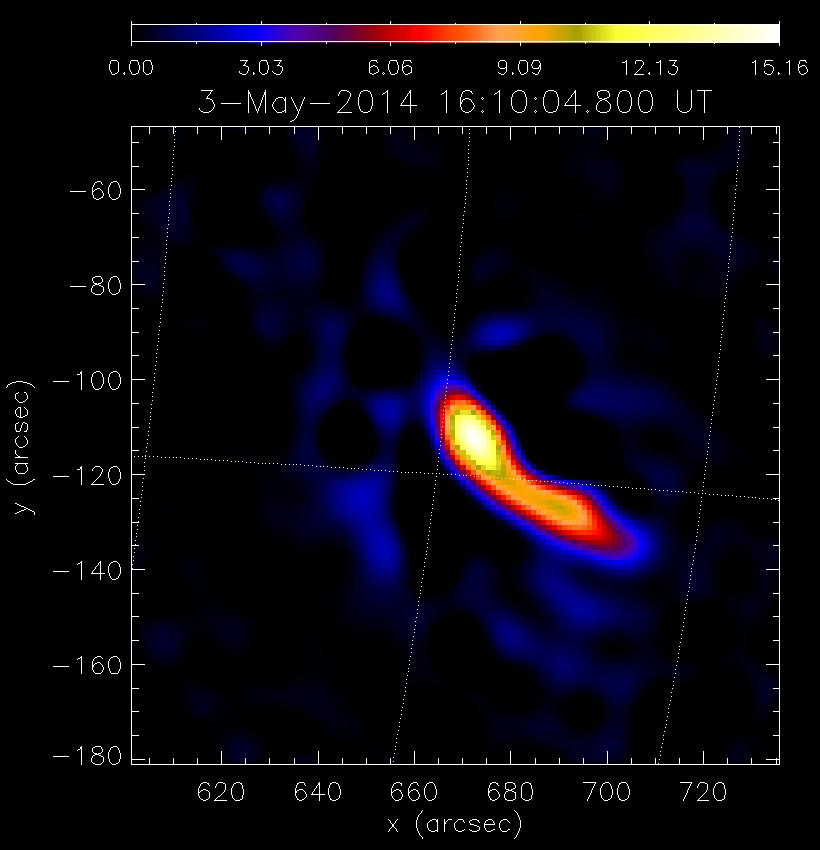}   
    \includegraphics[scale=0.11]{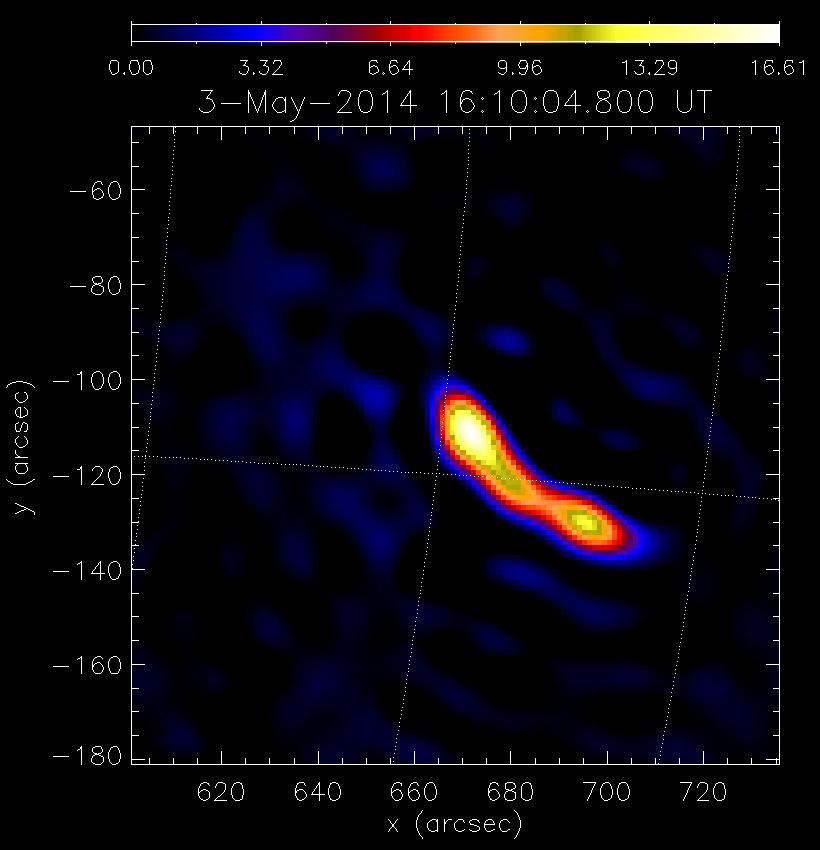} 
    \includegraphics[scale=0.11]{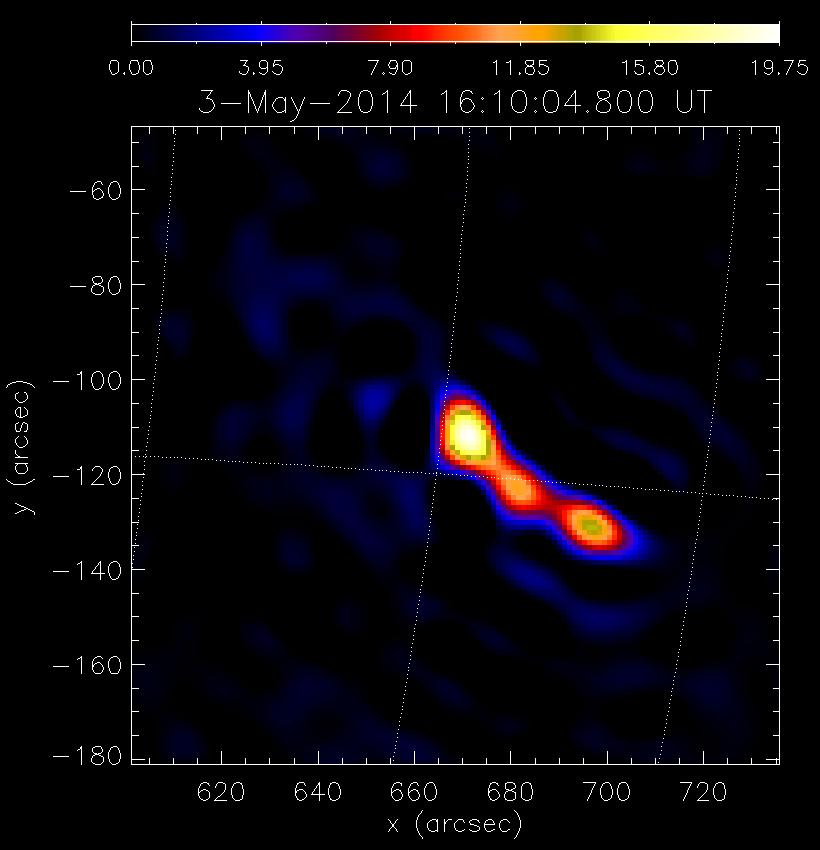} 
    \includegraphics[scale=0.11]{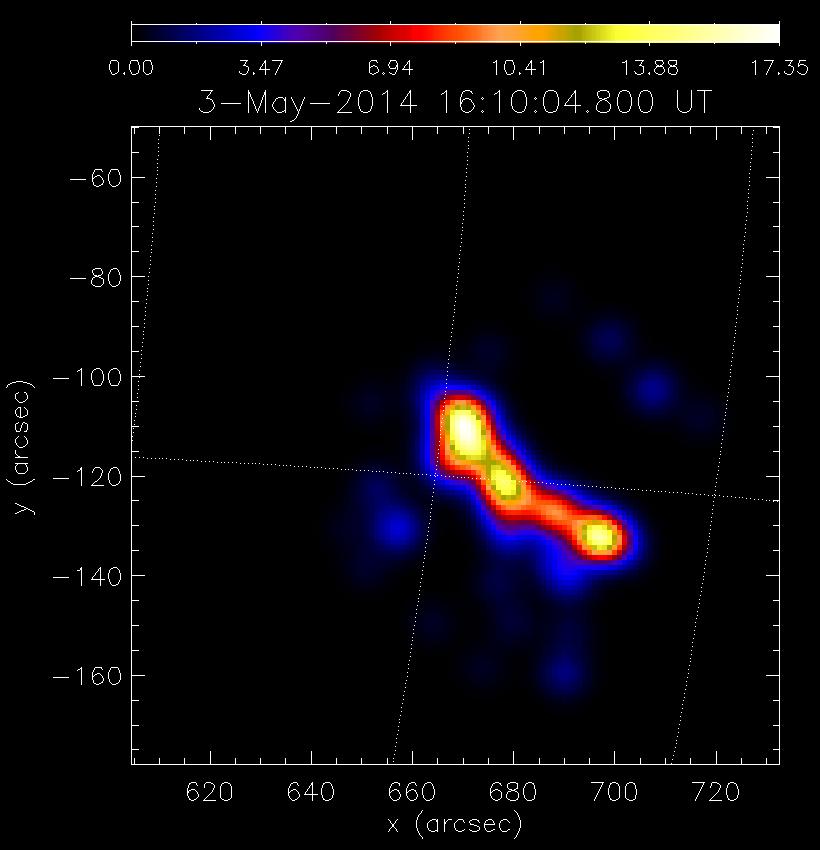} \vskip 0.1cm  
    \includegraphics[scale=0.11]{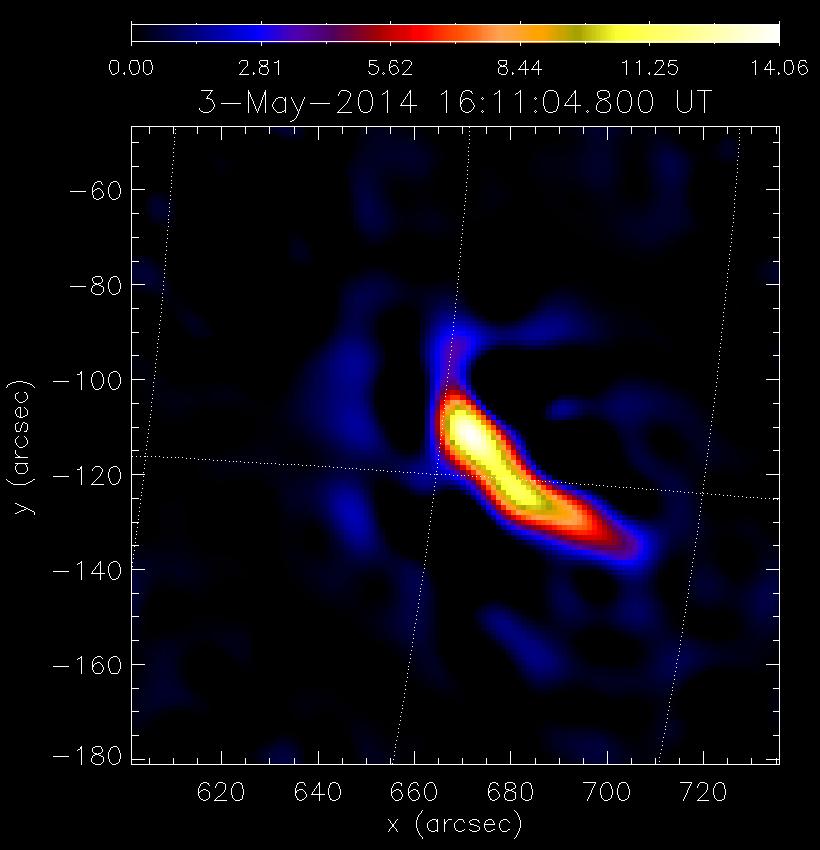}   
    \includegraphics[scale=0.11]{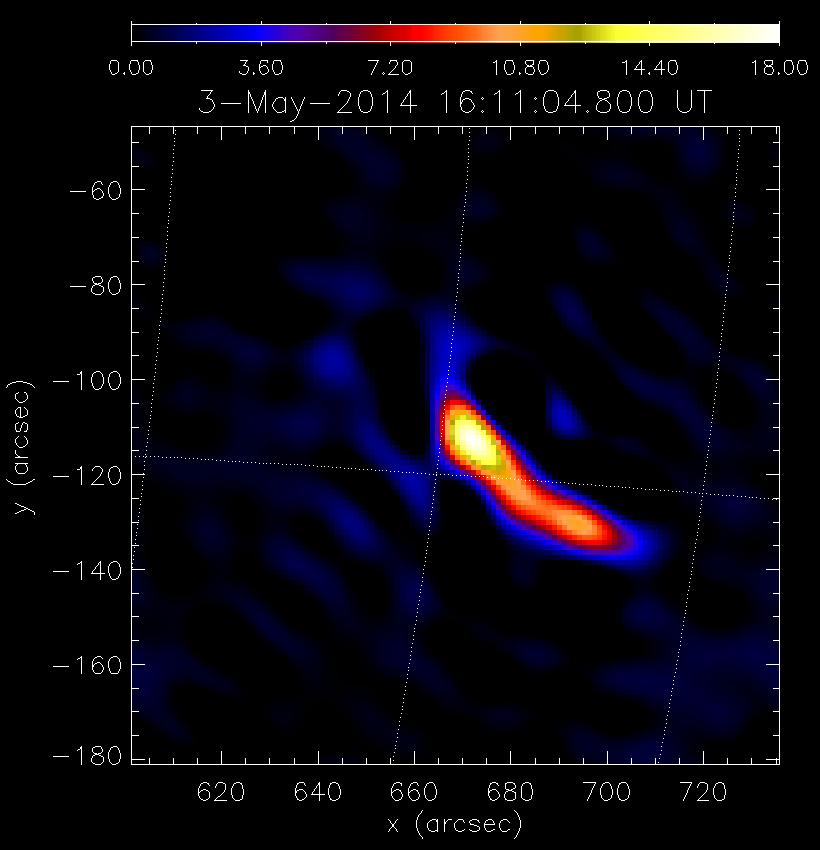} 
    \includegraphics[scale=0.11]{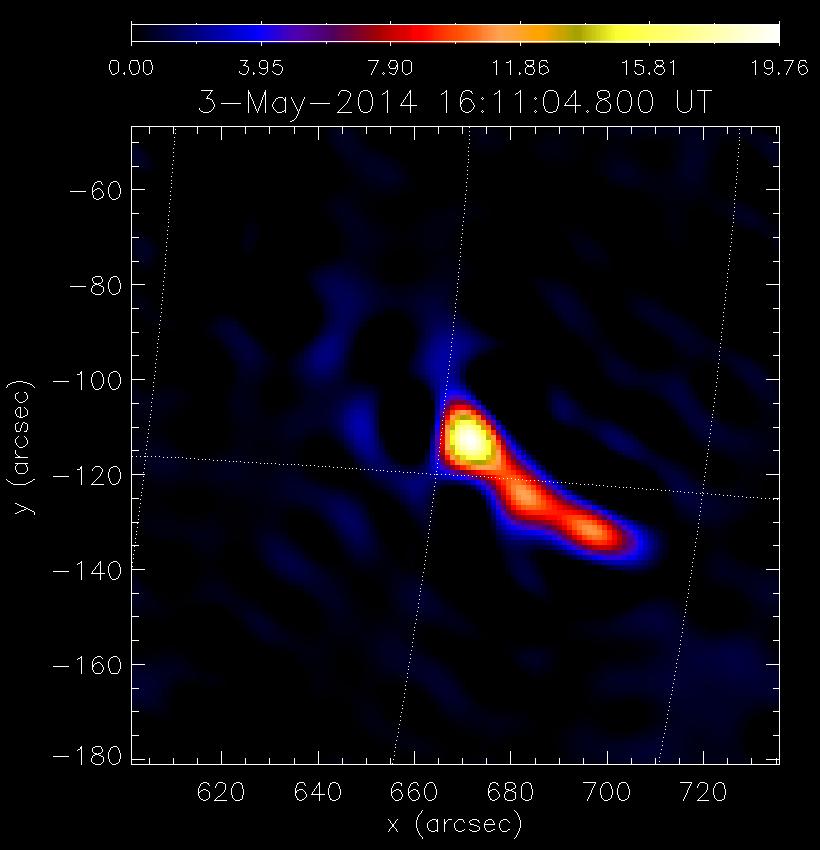} 
    \includegraphics[scale=0.11]{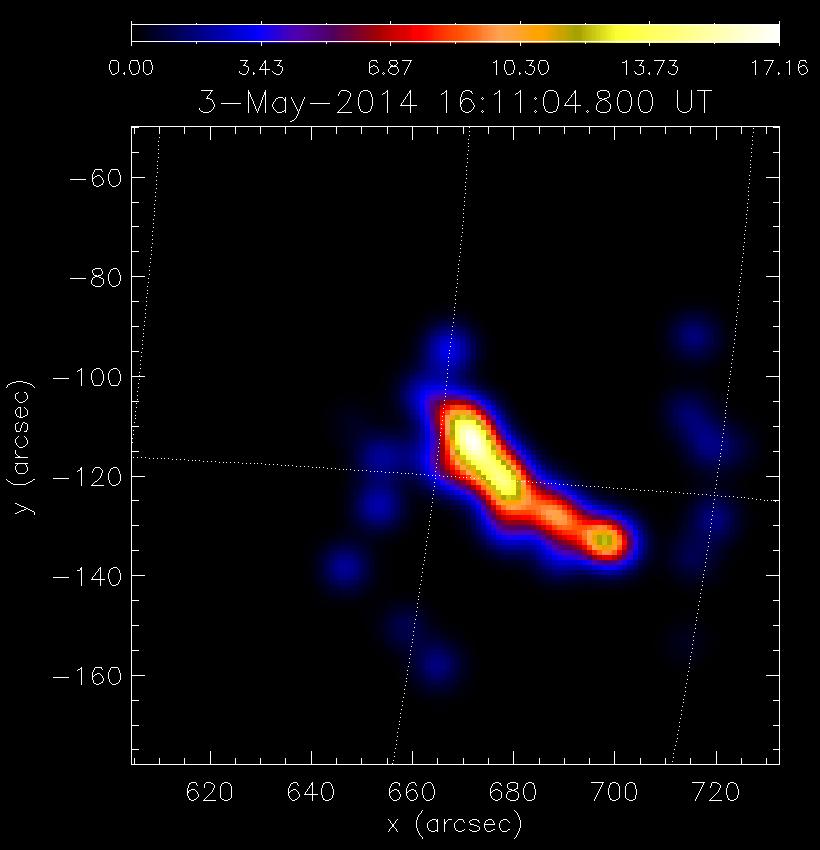}
\vskip 0.1cm  
    \includegraphics[scale=0.11]{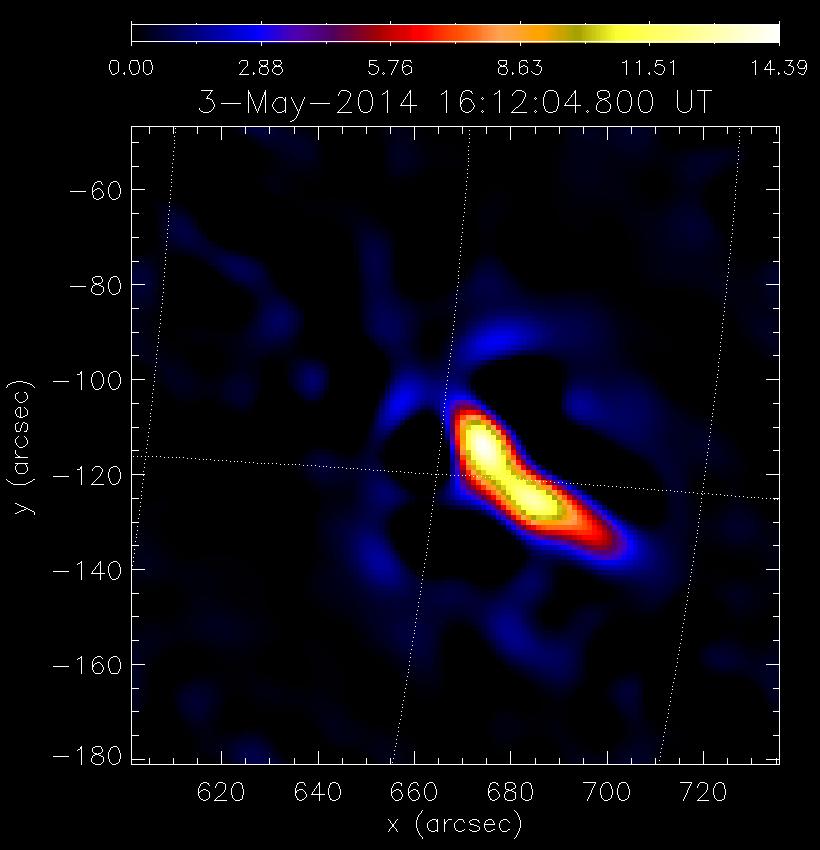}   
    \includegraphics[scale=0.11]{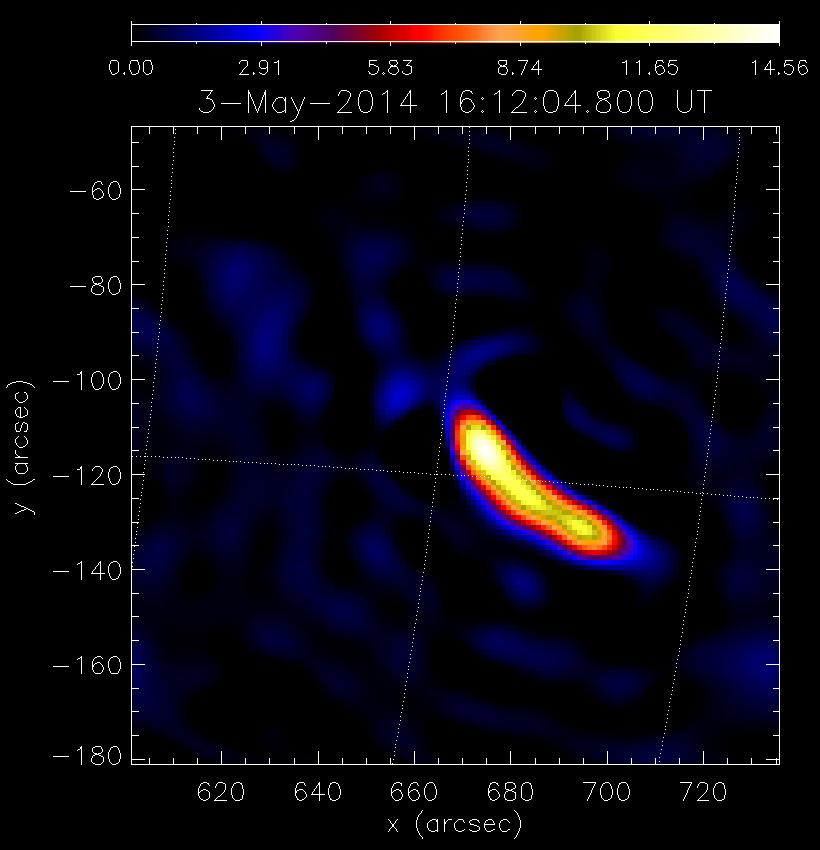} 
    \includegraphics[scale=0.11]{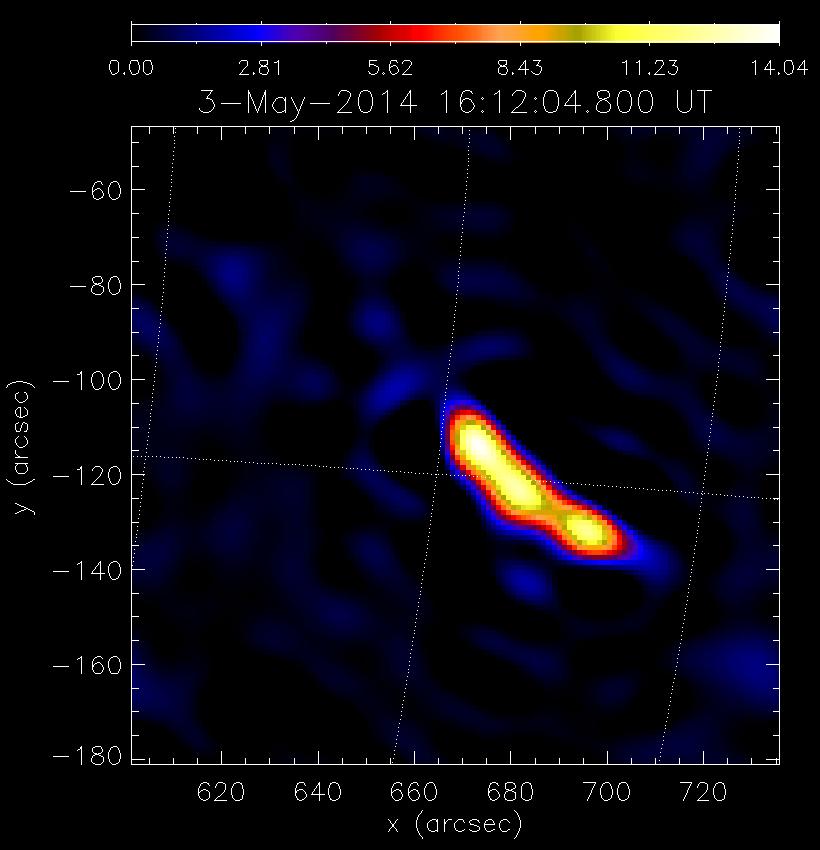} 
    \includegraphics[scale=0.11]{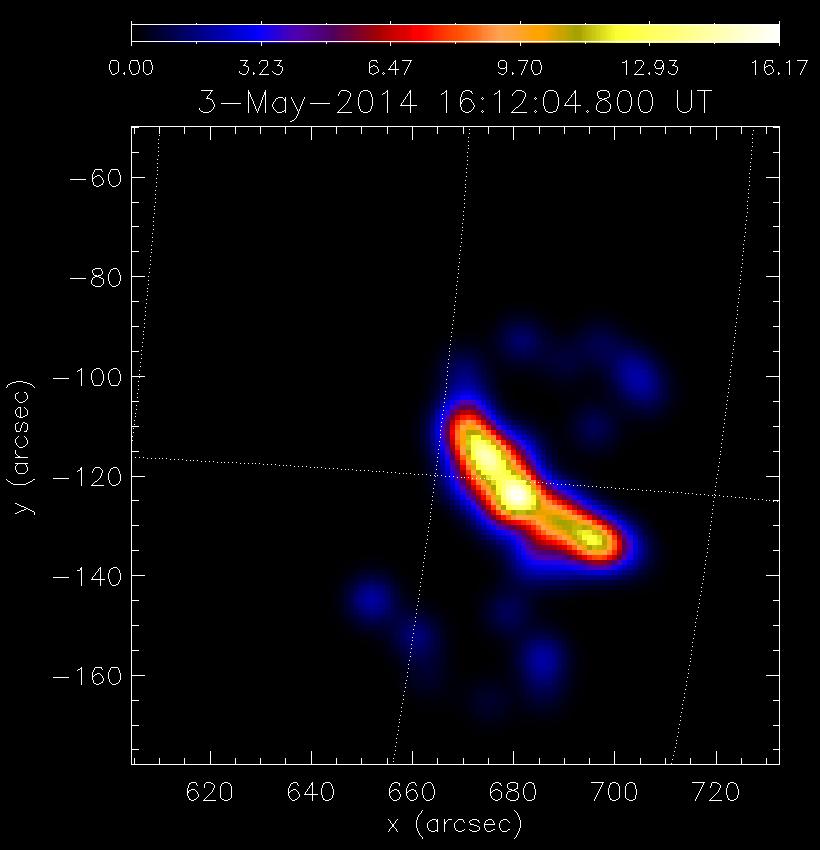}  \caption{Reconstruction of the flare observed by RHESSI on May 3 2014. From left to right, the columns contain the reconstructions obtained by uv$\_$smooth, uv$\_$smooth$\_$BP, uv$\_$smooth$\_$CC and CLEAN. From top to bottom, the rows denote the evolution of the flare shape in five time intervals from 16:08:04 through 16:12:04 UT (integration time: $1$ min).
}
    \label{fig_rhessi_may}
\end{figure}

\begin{figure}[h!]
\centering
     \includegraphics[scale=0.128]{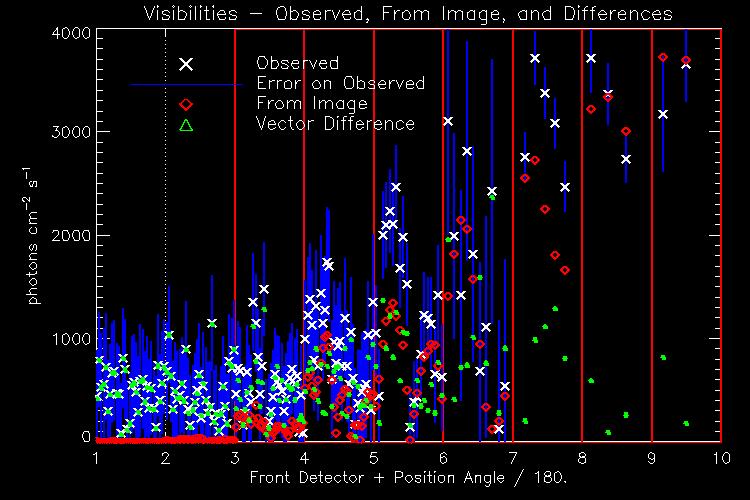}   
    \includegraphics[scale=0.128]{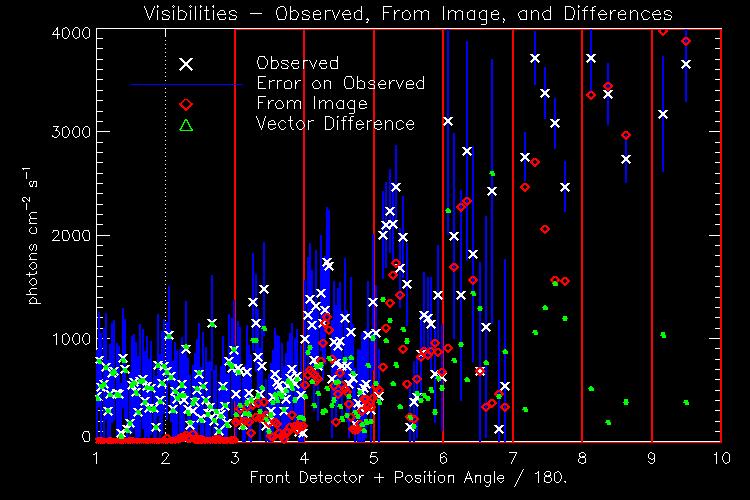} 
    \includegraphics[scale=0.128]{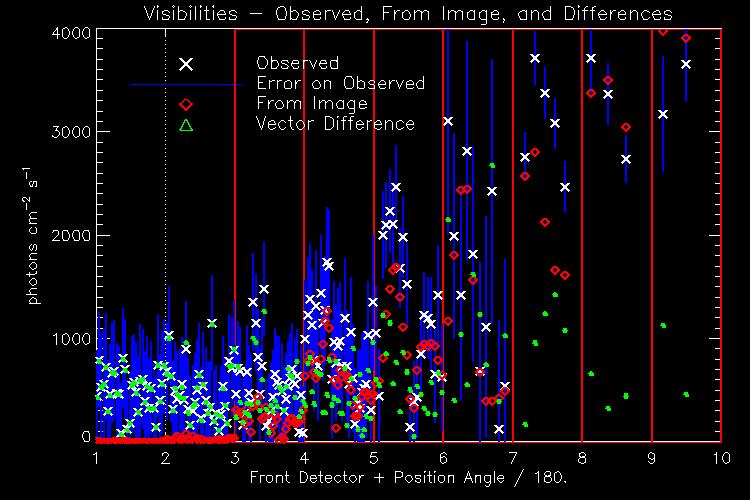} 
    \includegraphics[scale=0.128]{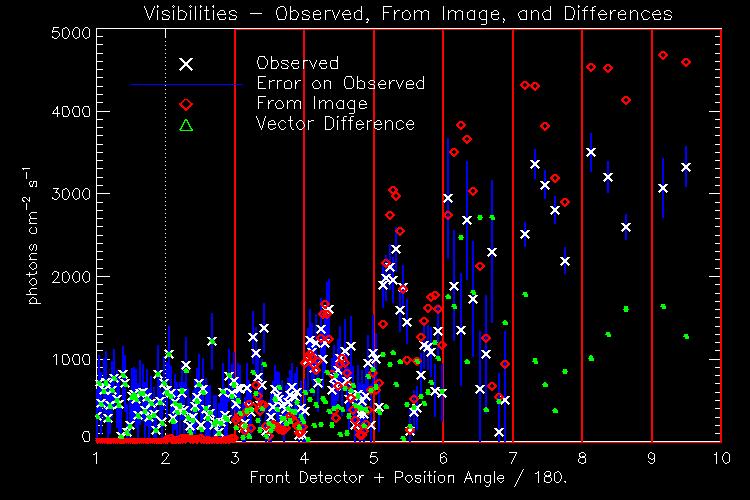}    
     \includegraphics[scale=0.128]{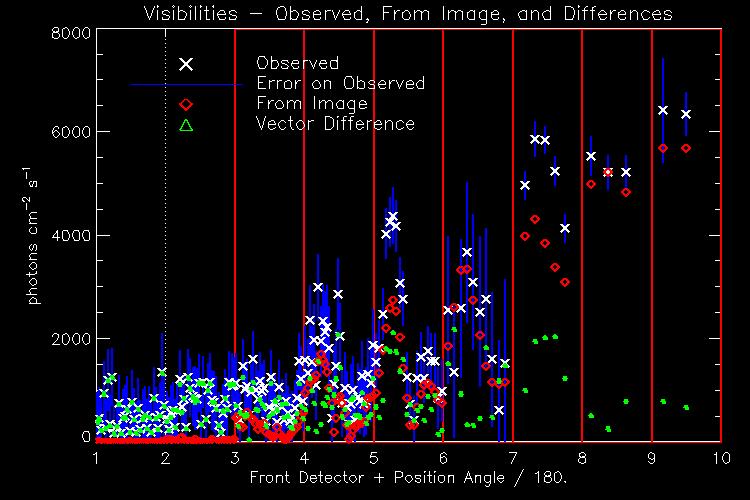}   
    \includegraphics[scale=0.128]{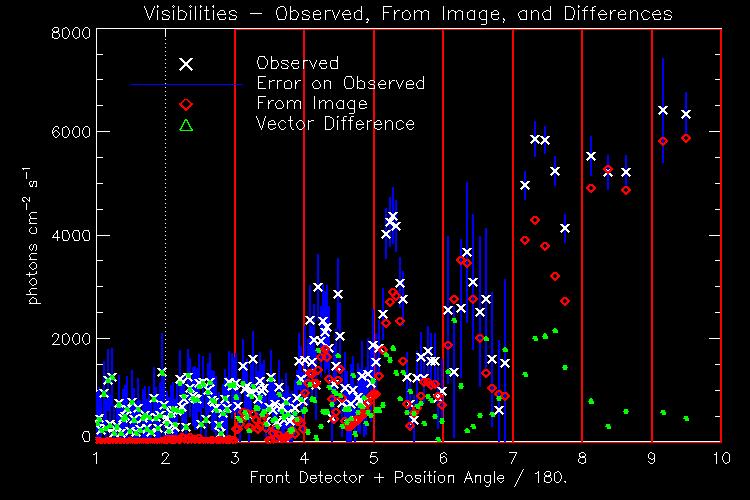}
    \includegraphics[scale=0.128]{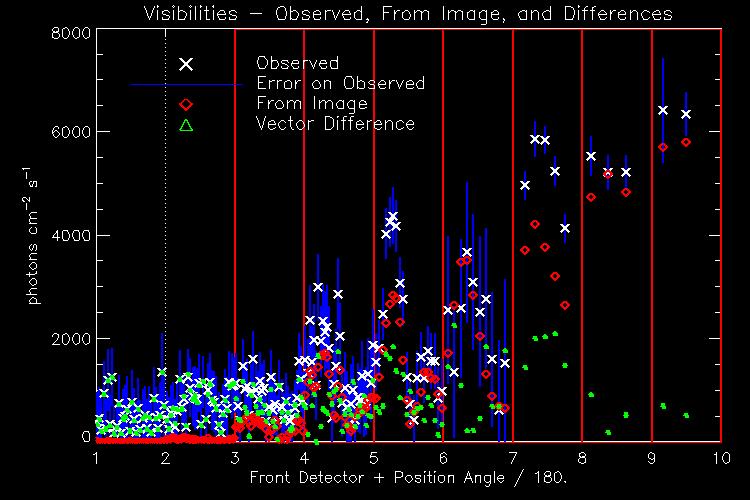} 
    \includegraphics[scale=0.128]{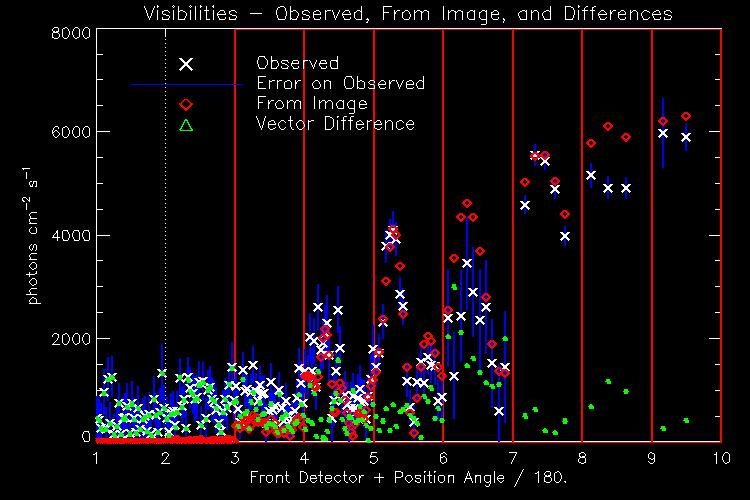} \vskip 0.1cm   
     \includegraphics[scale=0.128]{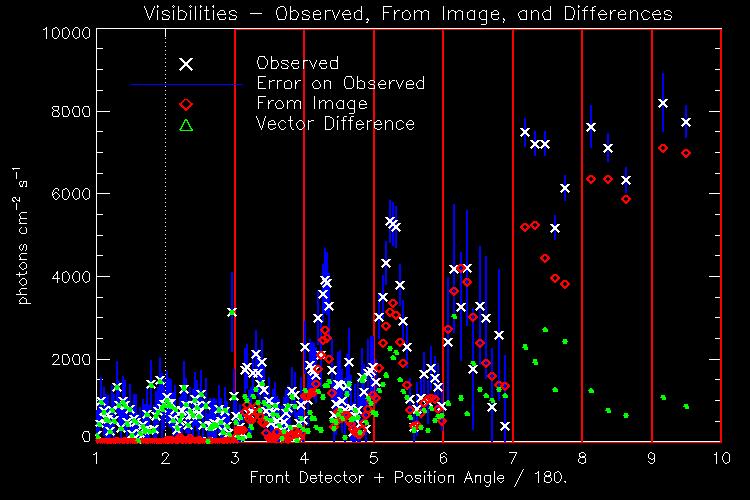}   
    \includegraphics[scale=0.128]{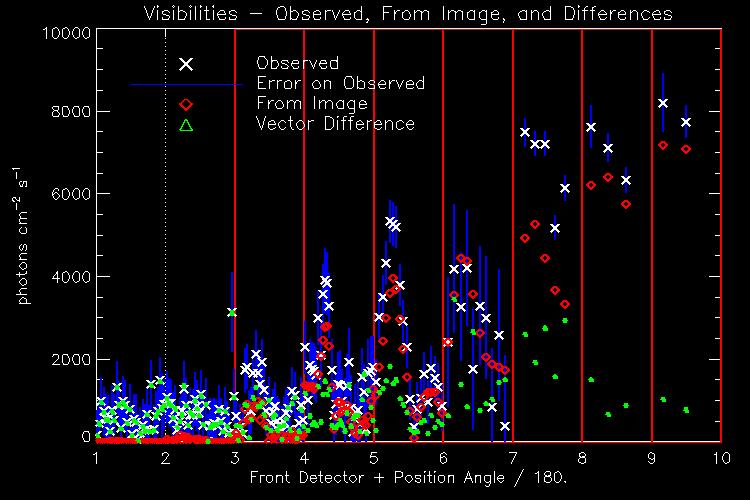} 
    \includegraphics[scale=0.128]{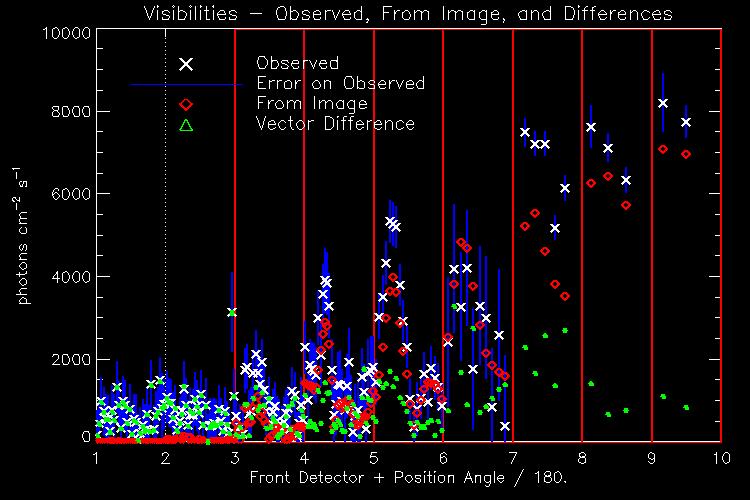} 
    \includegraphics[scale=0.128]{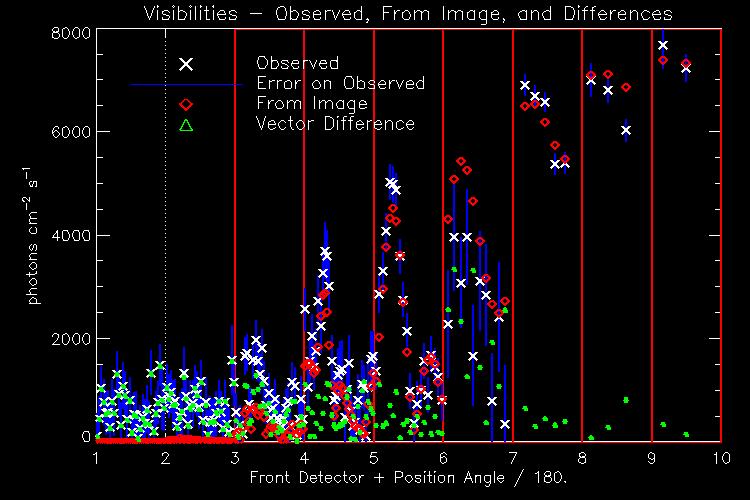} \vskip 0.1cm  
    \includegraphics[scale=0.128]{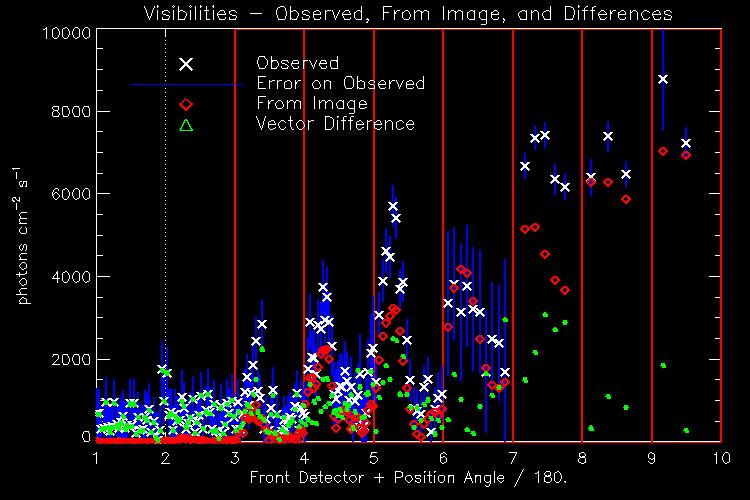}   
    \includegraphics[scale=0.128]{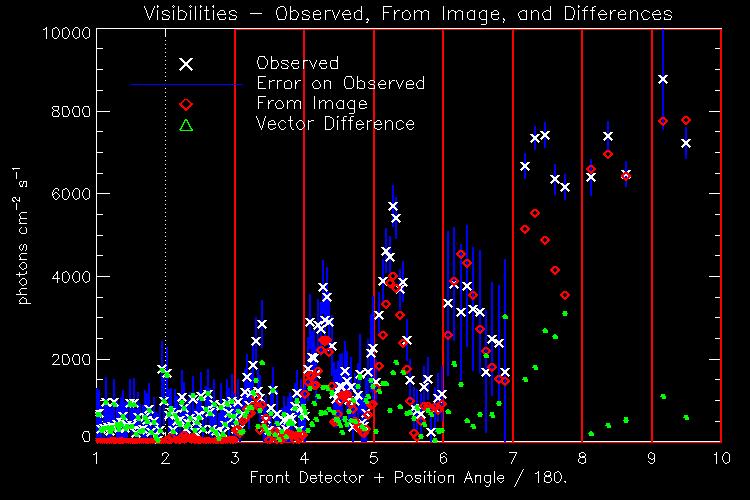}
    \includegraphics[scale=0.128]{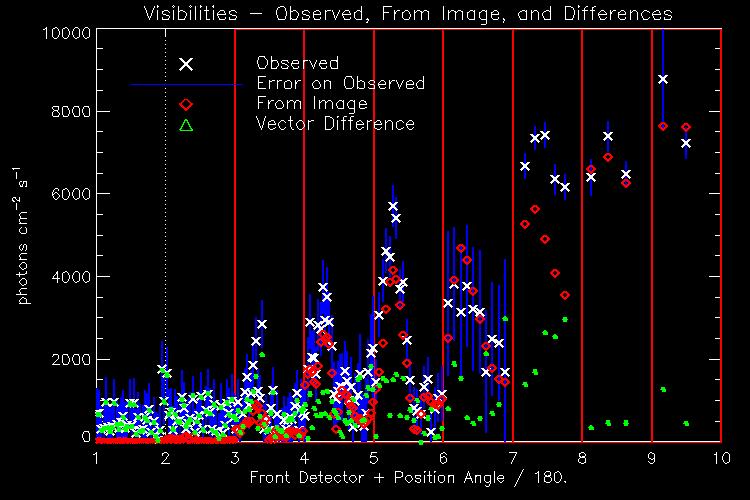}
    \includegraphics[scale=0.128]{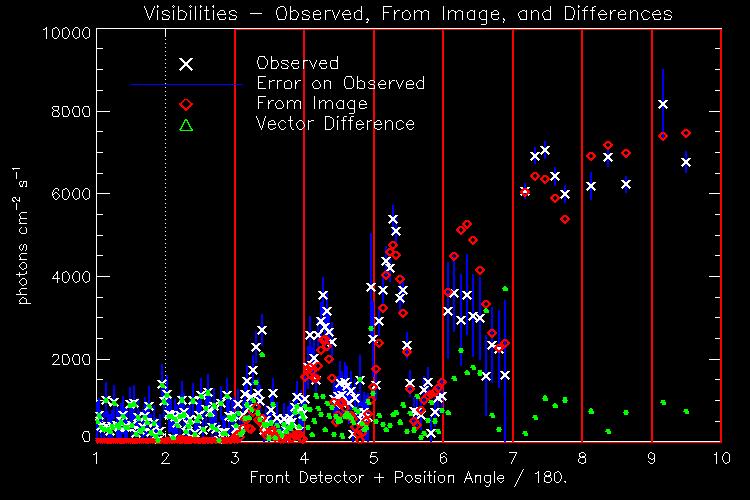}
\vskip 0.1cm  
    \includegraphics[scale=0.128]{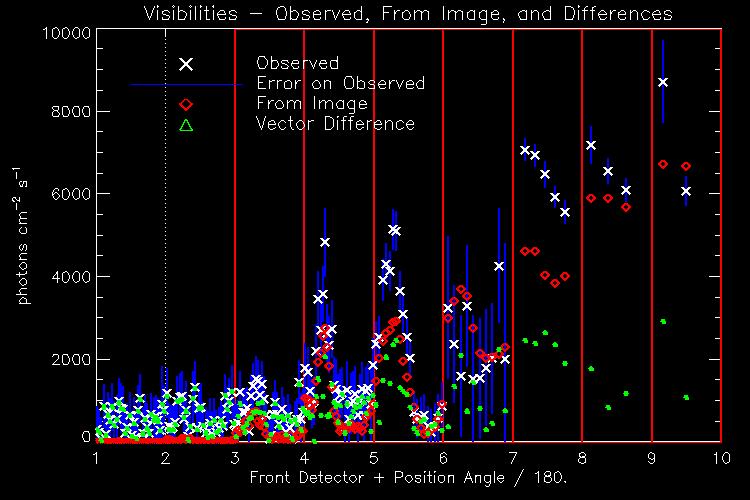}   
    \includegraphics[scale=0.128]{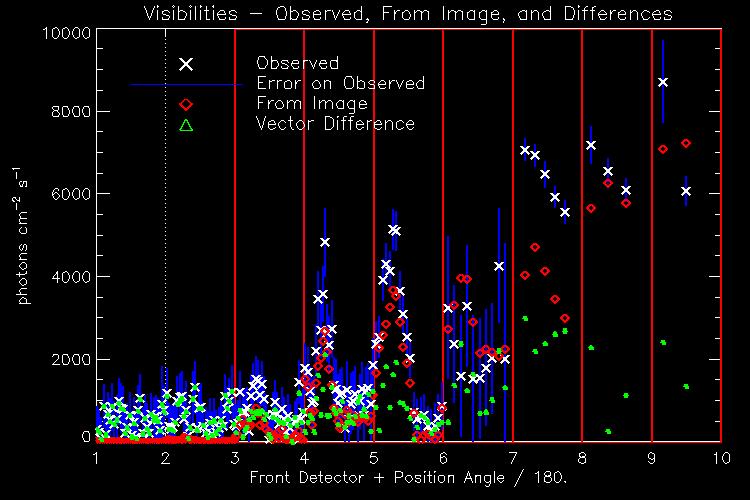}
    \includegraphics[scale=0.128]{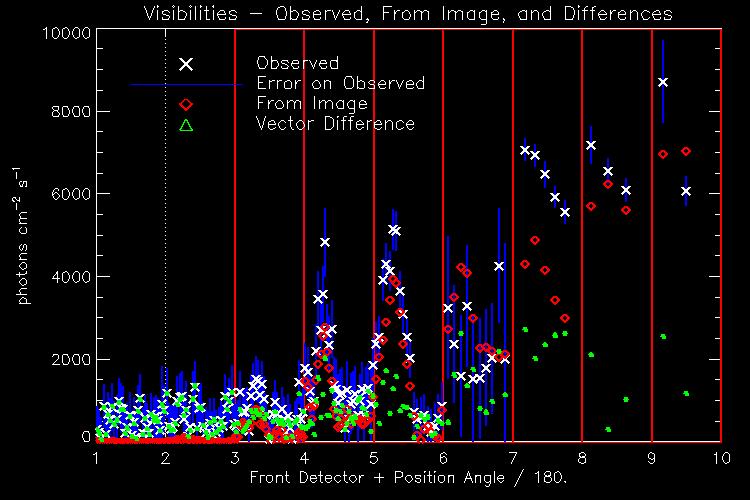} 
    \includegraphics[scale=0.128]{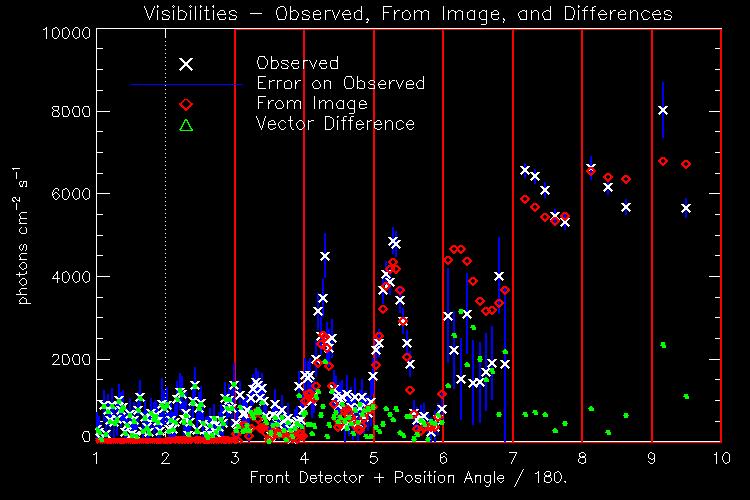}    
    \caption{Comparison between predicted and measured visibilities for the flare observed by RHESSI on May 3 2014. From left to right, the columns contain the fits corresponding to uv$\_$smooth, uv$\_$smooth$\_$BP, uv$\_$smooth$\_$CC and CLEAN. From top to bottom, the rows correspond to the evolution of the flare shape in five time intervals from 16:08:04 through 16:12:04 UT (integration time: $1$ min).    
}
    \label{fig_rhessi_may_vis}
\end{figure}

On Saturday, May 3 2014 the GOES $1-8$ $\mathring{A}$ passband instrument recorded nine C class flares originating from three different active regions. In particular, in the time interval between 15:54:00 UT and 16:13:40 UT {\em{RHESSI}} observed a C$1.7$ event whose flaring shape in the $3-6$ keV energy channel evolved from a double foot-point to a narrow ribbon-like configuration. We have tested the effectiveness of this enhanced approach to interpolation in the $(u,{\mbox{v}})$-plane by considering five time intervals in that range, each one of $1$ minute duration. First, we focused on the visibility bag recorded  at 16:07:04 UT by the combination of $3$ through $9$, $2$ through $9$ and $1$ through $9$ {\em{RHESSI}} detectors, respectively. Figures \ref{fig_rhessi_maydet} and \ref{fig_rhessi_vis_dets} respectively compare the reconstructions and the corresponding visibility fitting provided by uv$\_$smooth, uv$\_$smooth$\_$BP and uv$\_$smooth$\_$CC with the reconstructions and the fitting given by CLEAN when the map of the CLEAN components is convolved {\em{a posteriori}} with an idealized PSF with CLEAN beamwidth factor equal to $2$ (as done for the generation of the {\em{RHESSI}} image archive) and pixel dimension equal to $1$ arcsec in the 3 through 9 detector configuration and equal to $0.5$ arcsec for the other two combinations of detectors. The $\chi^2$ values of the four reconstruction methods are reported in Table \ref{tab:my_label_chi}. Then, in Figure \ref{fig_rhessi_may} and Figure \ref{fig_rhessi_may_vis} we fixed the configuration based on $3$ through $9$ detectors and compared the reconstructions provided by the same four imaging methods as in Figure \ref{fig_rhessi_maydet} and the corresponding fitting of the experimental measurements in the case of five time intervals between 16:08:04 and 16.12:04 UT. The $\chi^2$ values predicted by the four reconstruction methods with respect to the observations are contained in Table \ref{tab:my_label}.

\section{Comments and conclusions}
Enhancing visibility interpolation is particularly crucial in the case of the {\em{STIX}} image reconstruction problem, where observations are linked to a set of $30$ visibilities and, correspondingly, the sparsity of the sampling in the $(u,{\mbox{v}})$-plane is pronounced. As a confirmation of this, comparison with the four ground-truth configurations considered in the simulations of Figure \ref{figmap} shows that the use of VSKs provides more accurate estimates of the imaging parameters; this is particularly true in the case of Configurations 2 and 4 that produce wilder oscillations in the visibility domain and where the need of powerful interpolation is more urgent. The computational times reported in Table \ref{cpu} show that VSK interpolation increases the burden but keeps the reconstruction times competitive with the ones of most hard X-ray imaging methods. 

In the case of {\em{RHESSI}} observations, the use of finer grids increases the spatial resolution but, at the same time, introduces high resolution artifacts. However, also in this case we can notice an improvement carried by the use of VSKs with respect to standard uv$\_$smooth, i.e. the progressive fragmentation of the reconstructed sources is less significant particularly when detectors from $2$ through $9$ are used. For most cases, we can notice that uv$\_$smooth$\_$BP and uv$\_$smooth$\_$CC can guarantee a nice trade-off between reconstruction accuracy and fitting: imaging artifacts are less numerous and pronounced if compared to standard uv$\_$smooth while $\chi^2$ values are either comparable or smaller than the ones corresponding to CLEAN reconstructions. Further, comparison between uv$\_$smooth$\_$CC and CLEAN shows that the former method can be interpreted as a user-independent way to exploit the CLEAN component map. Therefore, uv$\_$smooth$\_$CC concludes the overall CLEAN process, keeping the highly reliable step providing the CLEAN components and replacing the more heuristic one represented by the convolution with an idealized PSF with a totally automatic process based on feature augmentation.

\begin{acknowledgements}
The authors acknowledge the financial contribution from the agreement ASI-INAF n.2018-16-HH.0. This research has been accomplished within Rete ITaliana di Approssimazione (RITA). This is the first paper that AMM and MP submit after Richard Schwartz passed away, on Saturday December 12 2020. In these difficult times for the whole humanity, Richard's death has represented a further reason of sadness and grief for the {\em{RHESSI}} and {\em{STIX}} communities. AMM and MP acknowledge that Richard's intellectual guide is and will always remain an unforgettable milestone for their current and future scientific activity.
\end{acknowledgements}

\begin{table}[ht]
    \begin{tabular}{ccccc}
    \hline 
    \hline
  detectors  & uv$\_$smooth & uv$\_$smooth$\_$BP & uv$\_$smooth$\_$CC & CLEAN \\
        \hline
         3--9 & 1.05 & 1.02 & 0.98 & 7.17 \\
         2--9 & 1.20 & 0.96 & 0.93 & 4.57 \\
         1--9 & 1.07 & 1.08 & 1.19 & 3.95 \\ 
    \hline 
    \hline
    \end{tabular}
    \caption{$\chi^2$ values predicted by the four reconstruction methods applied to the {\em{RHESSI}} visibilities observed on May 3 2014 at 16:07:04 UT. The values are computed with respect to the visibilities measured by detectors 3 through 9, 2 through 9, and 1 through 9, respectively.}
    \label{tab:my_label_chi}
\end{table}

\begin{table}[ht]
    \begin{tabular}{ccccc}
    \hline 
    \hline
    & uv$\_$smooth & uv$\_$smooth$\_$BP & uv$\_$smooth$\_$CC & CLEAN \\
        \hline
         $t_1$ & 1.13 & 1.10 & 1.07 & 3.70 \\
         $t_2$ & 1.80 & 1.73 & 1.75 & 1.71 \\
         $t_3$ & 2.25 & 2.18 & 1.96 & 1.42 \\ 
         $t_4$ & 2.75 & 2.14 & 2.06 & 1.90 \\  
         $t_5$ & 3.01 & 2.97 & 2.70 & 1.80\\
    \hline 
    \hline
    \end{tabular}
    \caption{$\chi^2$ values predicted by the four reconstruction methods applied to the {\em{RHESSI}} visibilities observed on May 3 2014 in the 5 time intervals from 16:08:04 through 16:12:04 UT (integration time: $1$ min). The values are computed with respect to the visibilities measured by detectors 3 through 9.}
    \label{tab:my_label}
\end{table}

\bibliographystyle{aa.bst}
\bibliography{bib_stix}


\end{document}